\documentclass[]{aa}
\usepackage{graphicx}
\usepackage{natbib}
\usepackage{multirow}
\usepackage{tabularx}
\usepackage{amssymb}
\usepackage{amsmath}
\usepackage{euscript}
\usepackage{bm}
\usepackage{pdflscape}
\usepackage{tikz}
\usepackage{pifont}
\usepackage{color}
\newcommand{\cmark}{\ding{51}}%
\newcommand{\xmark}{\ding{55}}%

\bibpunct{(}{)}{;}{a}{}{,} 

\newcommand*{\tran}{^{\mkern-1.5mu\mathsf{T}}}
\newcommand{\RHK}{$\langle R^\prime_{\rm HK}\rangle$}
\newcommand{\vect}[1]{\boldsymbol{\mathbf{#1}}}

\begin{document}

   \title{Estimating activity cycles with probabilistic methods II. \\ The Mount Wilson Ca H\&K data}
   \authorrunning{N. Olspert et al.}
   \titlerunning{Application of probabilistic methods to Mount Wilson Ca H\&K data}

   \author{N. Olspert \inst{1} \and 
   	J. Lehtinen \inst{2,1} \and 
   	M. J. K\"apyl\"a\ \inst{2,1} \and
    J. Pelt \inst{3} \and
       A. Grigorievskiy \inst{4}}
   \offprints{N. Olspert\\
          \email{nigul.olspert@aalto.fi}
	}
\institute{
ReSoLVE Centre of Excellence, Department of Computer Science, Aalto University, PO Box 15400, FI-00076 Aalto, Finland
\and Max-Planck-Institut f\"ur Sonnensystemforschung,
Justus-von-Liebig-Weg 3, D-37077 G\"ottingen, Germany
\and Tartu Observatory, 61602 T\~{o}ravere, Estonia
\and Department of Computer Science, Aalto University, PO Box 15400, FI-00076 Aalto, Finland
}

\date{Received / Accepted}

\abstract{
Debate over the existence of branches in the stellar
  activity-rotation diagrams continues. Application of modern time series
analysis tools to study the mean cycle periods in chromospheric activity index is lacking.}
{We develop such models, based on Gaussian processes, for one-dimensional time series and apply it to the extended Mount Wilson Ca H\&K
sample. Our main aim is to study how the previously commonly used assumption of strict 
harmonicity of the stellar cycles as well as handling of the linear trends affect the results.
}
{We introduce three methods of different complexity, starting with Bayesian harmonic regression model,
followed by Gaussian process (GP) regression models with periodic and quasi-periodic covariance functions.
We also incorporate a linear trend as one of 
  the components. 
  We construct rotation to magnetic cycle period ratio -- activity (RCRA) diagrams and apply a Gaussian mixture model to learn the optimal number of clusters explaining the data.
}
{We confirm the existence of two populations in the RCRA diagram; this finding is
  robust with all three methods used. We find only
  one significant trend in the inactive population, namely that the cycle periods get shorter with
  increasing rotation, leading to a positive slope in the RCRA diagram. This is in contrast
  with earlier studies, that postulate the existence of trends of different types in both 
  of the populations.
  Our data is consistent with only two activity
  branches (inactive, transitional) instead of three (inactive, active, transitional) such that
  the active branch merges together with the transitional one.
  The retrieved stellar cycles are uniformly distributed over the \RHK\, activity index, indicating that
  the operation of stellar large-scale dynamos carries smoothly over the Vaughan-Preston gap. At around
  the solar activity index, however, indications of a disruption in the cyclic dynamo action are seen.
}
{Our study shows that stellar cycle estimates from time series the length of which is short in comparison to the searched cycle itself depend
  significantly on the model applied.
  Such model-dependent aspects include the improper treatment of linear trends, while
  the assumption of strict harmonicity can result in the appearance
  of double cyclicities that seem more likely to be explained by the quasi-periodicity of the
  cycles.  
  In the case of quasi-periodic GP models, which we regard the most physically motivated ones,
  only 15 stars were found with statistically significant cycles against red noise model. 
  The periodicities found have to, therefore, be regarded as suggestive.
}

 
\maketitle

\section{Introduction}

Stellar cycles need to be studied to understand how stellar dynamos, including the solar one, operate.
It is clear that latitudinal and radial shear play important roles in the operation of the
solar dynamo, \citep[see e.g. the discussion in][]{BohmVitense2007}, but recent numerical
work of fully three-dimensional compressible magnetoconvection
in the regime of the solar \citep{Warnecke2017} and stellar \citep{Warnecke2017b} dynamos
has confirmed the
existence of various important turbulent effects that cannot be neglected.
These include the inductive effects arising from helical
turbulence, known as the $\alpha$-effect, and turbulent diffusion, known as the $\beta$-effect, 
that were reported to be active throughout the convection zone.
In particular, net advection of magnetic fields by turbulence, contributing to the effective
meridional flow, was found to be significant for the solar-like dynamo solutions \citep{Warnecke2017b}, indicating that turbulence has many effects
some of which have been neglected especially in the solar dynamo investigations so far. Therefore, even
the workings of the solar dynamo remain enigmatic.

From the dynamo modelling point-of-view, however,
it is crucial to know how the observed properties
of the dynamo depend on the basic system parameters (such as rotation). This provides
stronger constraints for dynamo models than the requirement of reproducing solely the
solar dynamo solution, which can be obtained by a wide variety of parameter combinations
and with very different dynamo models \citep[compare e.g.][]{KKT2006,Guerrero2008}.

One recent observational finding with far-reaching implications for dynamos was obtained
by \citet{WrightDrake2016}. They studied the X-ray luminosity of slowly rotating,
fully convective main-sequence stars, and compared the activity level with a sample
containing solar-like main-sequence stars.
The former class of stars cannot possess layers of strong shear in the transition region
between the radiation and convection zones, called tachocline. The latter class,
to which the Sun belongs, will most likely exhibit such layers, verified by
helioseismology for the Sun. Both classes were concluded to exhibit very
similar levels of magnetic activity, and its dependence on
rotation is also nearly identical. Together with the recent modelling results
discussed above, highlighting the importance of turbulent processes, these findings
cast doubt on the role of the tachoclinic shear layers alone being decisive for stellar dynamos.

Crucial sources of information to determine stellar cycles are long-term data sets monitoring the magnetic activity
level and its variability in stars. To detect a solar-like cycle, however, such monitoring
should be performed over decades, narrowing down the number of useful datasets to a very few.
Such data sets are available in photometry,
for example the Tennessee State University T3 photometry since 1987 \citep{Henry1995T3},
the Vienna photometry since 1991 \citep{Strassmeier1997ViennaAPT}, and the
ASAS photometry since 1996 \citep{Pojmanski1997ASAS},
as well as in chromospheric line emission. Of the latter,
the Mount Wilson (MW) chromospheric Ca H\&K sample has been one of the main
datasets to study stellar activity cycles for decades, attracting attention
up to the present day
\citep[see e.g.][]{Noyes1984,Baliunas1995,BST98,SB99,BohmVitense2007,Olah2016,Metcalfe2016,Brandenburg2017,BS18}.

A basis for many of these studies, concentrating on the dependence of stellar cycles on the stellar parameters,
is the magnetic period determination undertaken in \citet{Baliunas1995} (hereafter Cyc95) using the Lomb-Scargle (LS)
method. While the aim in Cyc95 was to detect periods stable over the whole data set length,
\citet{Olah2016} used short-term Fourier transform to detect locally active periods. 
Many studies have attempted to find dependencies of the cycle length or the ratio of cycle length
to rotation period as function of some stellar activity measure; in this study we concentrate
on the latter type of dependency, which we label as RCRA (Rotation period-to-Cycle length Ratio -- Activity).
By using these quantities one aims at removing the dependence of the cycle length on the poorly known
convective turnover time $\tau$, as originally pointed out by \citet{Tuominen1988}. Many studies,
however, bring back this dependence by considering the Rossby (or Coriolis) number instead of an independent
activity measure \citep[e.g.][]{SB99}. As the activity indices themselves are correlated
with rotation, some studies rely on even simpler statistics by plotting cycle length vs.
the rotation period \citep[e.g.][]{BohmVitense2007}; we have also studied these diagrams and
denote them with CR (Cycle length -- Rotation period) diagrams.

\citet{BST98} were the first to report on the existence of two different linear branches
with positive slopes in RCRA diagram, dividing the stars into two groups depending on their
activity index, based on the Cyc95 period analysis. A third branch of
super-active stars with a negative slope was identified by \citet{SB99},
by combining the cycle length estimates from the MW sample with estimates from photometric and other 
studies. They had to resort to diagrams with Rossby numbers on the
abscissa due to the non-existence of chromospheric activity indices for the super-active stars.
\citet{BohmVitense2007} found positive linear scalings in the CR diagram of the Cyc95 data,
and observed that the Sun is a peculiar case lying in between the branches.
Similarly, \citet{Metcalfe2017} studied data from various sources, including MW and Kepler
targets and showed two different linear relationships in a CR diagram.
By analysing the extended MW sample using a Generalized Lomb-Scargle method (GLS),
\cite{BS18} no longer find the Sun being an outlier, and claim that the active branch is
caused by a selection effect arising from the upper and lower limit of cycle lengths.

These positive
`detections' of branches are accompanied with many `non-detections' even when
the same data is used \citep[compare e.g. the results of][]{Baliunas1996,BST98}.
Most often the negative detections, however, come from different datasets.
The other sources that have long enough time extents are photometric time series. If
long enough, usually results consistent with the MW sample are obtained \citep[see e.g.][]{Lehtinen2016}.
However, in \citet{Distefano2017}, where a photometric sample of 49 near-by main-sequence (MS) stars
from the ASAS campaign with the extent of roughly ten years
were analysed, no correlation between cycle length and rotation period was found.
This has been explained by 
the hypothesis that these stars are mostly belonging to the transitional region between the active and
superactive branches.
Neither was in the 
study of Kepler stars by \citet{Reinhold2017} detected any functional dependence
between 
these quantities.
This was in turn explained by the hypothesis that 
on the active
and superactive branches the stars show less signatures in their
photosphere thus not detected using Kepler data.

The aim of this study is to re-analyse the recently released extended MW data set with different kinds of
probabilistic period search methods of increasing complexity. We investigate the reliability of the earlier determinations
with a strictly harmonic model, and the overall limitations related to retrieving cycle lengths from data sets the extent
of which is in the same order of magnitude as the desired cycles.
We start the cycle search with a Bayesian generalised Lomb-Scargle periodogram with trend\footnote{We will interchangeably refer to it as 
a harmonic model, to emphasize its distinction from periodic and quasi-periodic models.} 
(BGLST) introduced in \citet{Olspert2017a} (Paper I).
Then we explain the necessity of including the trend 
component directly into the model as opposed to detrending the data beforehand or leaving the data undetrended. 
Next we apply more complex, periodic and quasi-periodic Gaussian 
process (GP) models to the data, construct the RCRA and CR diagrams 
for the subset of stars
with known rotational period, and as the last step, we apply the Gaussian mixture model (GMM) to it. 
We show that two activity clusters describe the data better than one or three.

In the next section we shortly describe the MW dataset and how we pre-processed it before starting the analysis. 
After that we will discuss the points which led us to the simplest model, that is, the harmonic model with trend. 
Subsequently we briefly discuss the issue of rotational period removal from the data and then turn to the 
description of the regression models. 
In the end we discuss the significance of the results to the theory.

\section{Data and pre-processing}\label{data}
In the current study we analyse the dataset of one of the chromospheric activity proxies known as MW Observatory 
S-index\footnote{\url{ftp://solis.nso.edu/MountWilson_HK/}}. Its definition can be found for example in \citet{Egeland2017}. The MW dataset being recently made publicly available contains time series of more than 2300 objects
including the Sun, 137 of which were monitored for more than ten years,
and are therefore suitable to be used in this study.
The time span of the observations is ranging for most of the stars from years 1966 to 1995 while for 35 of the stars up to the year 2001. 

For some stars, however, the full range of observations was unsuitable for use.
Namely, a closer look at the raw data 
revealed some possibly erroneous 
measurements, including full observing seasons potentially biased and other abrupt systematic changes in the data. To eliminate these problems 
we pre-processed the time series of 13 stars by omitting the corresponding measurements. 
From all the time series we further omitted all data points residing farther than $4\sigma$ distance 
from the sample mean.
In the case of the Sun we used a hybrid dataset combined from the MW S-index dataset and
the Sacramento Peak Ca K observations transformed into the same scale as the
MW S-index. We used the scaling $S_{\rm Sac} = 2.61K_{\rm Sac} - 0.0647$
for the Sacramento Peak data, which we determined as a linear fit onto coeval
solar observations from MW and Sacramento Peak.

To make the computations feasible for GP models, we further downsampled the datasets
to keep only one data point per every ten days from the original dataset.
We tested that this had negligible effect on the results, which is expected as we are estimating cycle periods starting from 2 yrs and up.
For the harmonic model, however, no downsampling was used.

\section{Finding the lengths of the activity cycles}\label{methods}
In the current work we use the so called effective methods to estimate the cycle lengths. Such methods
do not solve for any physical model directly, but describe the behaviour of the system with functions
that depend on a certain set of parameters, that in turn have a relation to the physical quantities, such as
rotation period or activity cycle length of a star.

First thing to notice is that the time series of stellar activity proxies cannot be assumed to be strictly periodic. 
The most clear evidence of deviation from pure periodicity can be seen for instance in the 11 year cycle of the Sun.
This is commonly interpreted as the solar dynamo behaving in a similar manner as a nonlinear
oscillator in a quasi-periodic regime. This can be verified when solving for the dynamics together with the dynamo equation,
when the frequency and amplitude become correlated (Waldmeier rule), 
and also the rise and fall times of the cycle
are different \citep[for one of the earliest examples, see][]{Yoshimura78}. The stellar dynamos have no reason
to be any less nonlinear -- on the contrary, the nonlinearities can be expected to be stronger in more
active stars.
From this perspective, for modelling the datasets, it would be preferable to use quasi-periodic models over the periodic 
ones, however
the stellar activity datasets
currently available span maximally several decades, limiting the count of observed cycles to 
very low numbers.
It opens the question whether fitting the more complex GP models is feasible to the given data or should one restrict to a harmonic model. 
We will discuss the related issues in Sect.~\ref{GP_models} and get some
empirical evidence of usefulness of the quasi-periodic GP model in Sect.~\ref{GP_QP_tests}.

As an alternative to the GP methods we mention yet another 
effective method developed for cycle search, namely the $D^2$ phase dispersion statistic \citep{Pelt1983}. When used with variable coherence length 
this method is suitable for time series with slowly changing period or modulated phase and amplitude. We have previously applied this method to estimate 
the rotational period of the young solar analogue LQ Hya \citep{Olspert2015} and magnetic activity cycles in multidimensional magnetohydrodynamic 
simulation data \citep{Olspert2016}. We did not consider using the given method here though, as the $D^2$ statistic 
in the given formulation would not allow us to incorporate a trend. 
We will, however, want to avoid detrending the data beforehand due to the possible issues discussed below.

Different past studies have handled the MW dataset slightly differently. In Cyc95 the cycle lengths 
were calculated using the LS
periodogram \citep{Lomb1976, Scargle1982} with detrending in case of need, while
in \citet{Olah2016} time-frequency analysis methods were used to extract locally persistent periods from the data.
In the latter study only the stars with long and continuous observations (no gaps in the seasons) from the MW sample 
were used. The selection criteria remain fully unclear though, as the majority of these stars have well determined cycle
periods from Cyc95. 
The methods in these two studies contrast significantly as Cyc95
is primarily dedicated to analysing stationary, but \citet{Olah2016} to nonstationary time series, that is, finding 
globally persistent vs. local periods respectively.

Our aim is to focus on an approach similar to Cyc95 and we start the analysis with the method as simple as possible. Most obvious 
choice for that would be a harmonic model similar to GLS
periodogram \citep{Zechmeister2009}, as was used in the study of \cite{BS18}.
However, closer look at the MW data reveals apparent trends in case of many stars in the sample. This 
opens up the question whether we should remove the linear trend before the period search or leave the data undetrended.
In Paper I we address this question more thoroughly and give empirical evidence that similarly to why the GLS
method is preferred over the approach consisting of first centring the data and then applying 
the LS method, it is advantageous to directly include linear trend component into the regression 
model rather than doing detrending
followed by application of any period search method. 
The opposite also holds -- leaving the data undetrended before doing the period search is generally less beneficial, 
compared to the model including the trend component. 

Returning to the discussion of the MW dataset, it is hard to say if the apparent trends are real, caused for example by instrumental 
effects, or are they actually manifestations of very long cycles or even both. 
While the last explanation may seem more likely as the trends vary from dataset to dataset,
answering this question fully is difficult and, in fact, is not strictly necessary.
Instead we find a workaround to this problem by using the BGLST method.
This way by optimizing the parameters we assume that 
the best estimate for the trend will be automatically inferred from the data.
Before moving to the regression models used in cycle detection we add a couple of 
remarks about removing the rotational signal from the data.

\subsection{Removing the rotational signal}\label{removing_rot}
The estimation of the rotational period being a complicated task enough is not part of the current study and
we leave the details of it for an upcoming paper. Here we just mention that the estimates were obtained by fitting the 
periodic GP to individual observing seasons and refining these estimates using the Continuous Period Search (CPS) method 
\citep{Lehtinen2011}.
This procedure gave us robust error estimates for the rotation periods and allowed
us to assess their reliability. We only accepted period estimates from
those stars where the CPS produced reliable fits that agreed
with the GP periods.
For a handful of objects, the rotation periods were refined w.r.t. previous studies, and for a few
stars with no previous period detections, rotation periods were obtained.
For the current study these refinements are of minor importance, and therefore
not discussed here any further.

Before the activity cycle search, we removed the rotational signal from the data
using the following procedure:
We first removed the 
slow trends from the data by fitting a GP with 
squared-exponential kernel with a time-scale\footnote{In GP literature this parameter is generally called length-scale, but in the context of time series we find the term time-scale to be more natural.} of one year. 
Then we calculated GLS periodograms and performed spectral cleaning in the narrow band of frequencies around the 
known rotational period using 99\% significance level (more about cleaning in the next section). Finally we added 
back the slow trend model. The calculations showed that the reduction of variance in most cases was only couple of 
per cents, but in the best case as high as 39\% for the stars HD68290 and HD32008.

\subsection{BGLST model}\label{harmonic_model}
As the simplest case we decided to use the BGLST method (for more detailed discussion see Paper I),
which is defined as the following regression model:
\begin{equation}\label{eq_harmonic_model}
y(t_i)=A\cos(2\pi f t_i - \phi) + B\sin(2\pi f t_i - \phi) + \alpha t_i + \beta + \epsilon(t_i),
\end{equation}
where $y(t_i)$ and $\epsilon(t_i)$ are the observation and noise at time $t_i$, $f=1/P$ is 
the frequency of the cycle, $A$, $B$, $\alpha$, $\beta$ 
are the parameters to be optimized but $\phi$ is set to a frequency dependent value 
such that the orthogonality of the $\cos$ and $\sin$ functions on a nonuniform set of $t_i$ is guaranteed
\citep[see][]{Mortier2015}.
We assume 
that the noise is Gaussian and independent between any two time moments, but we do not 
assume the constancy of variances. For parameter inference we use the Bayesian model, 
where the marginal posterior probability for frequency $f$ is given by 
\begin{equation}\label{eq_posterior_marginal}
p(f|D) \propto \int p(D|f,\vect{\theta})\mathcal{N}(\vect{\theta}|\vect{\mu}_{\theta}, \vect{\Sigma}_{\theta}) d\vect{\theta},
\end{equation}
where $p(D|f,\vect{\theta})$ is the likelihood of the data, 
$\vect{\theta} = \left[A, B, \alpha, \beta\right]\tran$ 
and $\mathcal{N}(\vect{\theta}|\vect{\mu}_{\theta},\vect{\Sigma}_{\theta})$ is their Gaussian independent prior distribution. 
For the frequency $f$ we use a uniform prior. 
The likelihood is given by
\begin{equation}\label{eq_likelihood}
p(D|f,\vect{\theta})=\left(\prod_{i=1}^{N}\frac{1}{\sqrt{2\pi}\sigma_i}\right)\exp\left(-\frac{1}{2}\sum_{i=1}^{N}\frac{\epsilon_i^2}{\sigma_i^2}\right),
\end{equation}
where $N$ is the number of data points, $\epsilon_i=\epsilon(t_i)$ and $\sigma^2_i$ 
is the noise variance at $t_i$.

The derivation of a closed formula for the spectrum Eq.~(\ref{eq_posterior_marginal}),
the selection of priors for $\vect{\theta}$,
the error and significance estimation of the optimal period are discussed in more 
detail in Paper I. 
Here we only make some comments about particular 
approaches used in the current study.
We followed the spectral cleaning procedure similar to 
\citet{Roberts1987} until no more significant peaks were found
and we claim significant all the peaks which satisfy the following quite strong criteria:
\begin{equation}\label{delta_bic}
\Delta {\rm BIC}={\rm BIC}_{M_{\rm null}} - {\rm BIC}_{M_{\rm H}} \ge 6,
\end{equation}
where $M_{\rm null}$ is the linear model without harmonic $M_{\rm H}$ is the BGLST model, 
${\rm BIC}=\ln(n)k-2ln(\hat{L})$ is the Bayesian Information Criterion (BIC), $n$ is the number of data points, $k$ the number of model parameters and $\hat{L}=p(x|\hat{\theta},M)$ is the likelihood of data for model $M$ using the parameter values that maximize the likelihood. 

Due to the nonuniformity in the average sampling frequency in the data 
(in most cases the subset of data spanning until the year 1980 is much more sparse 
compared to the later subset), the values of ${\rm BIC}$ were not calculated for the 
original data, but instead for the seasonal means. This way we reject the models whose
harmonic parts fit well only those regions of the data, which are more densely 
populated (moreover we avoid finding only locally stable periods).

When all the harmonics are extracted, we calculate once more the spectra separately for each of them 
and assume them to represent true probability distributions.
One can then fit the Gaussian to the spectral line to estimate the error of the frequency 
\citep[Chapter~2.4]{Bretthorst1988}. This technique is known as Laplace approximation.
However, we take the simpler path and calculate the proxy for the error estimate as 
$\sigma_f^2=\mathbb{E}[(f-f_{\rm opt})^2]=\int (f-f_{\rm opt})^2 p(f|D,M_{\rm H})df$, 
where $f_{\rm opt}$ is the optimal frequency found for the harmonic\footnote{To get the 
proper posterior $p(f|D,M_{\rm H})$ we need to normalize the value obtained from 
Eq.~(\ref{eq_posterior_marginal}) by an integral over sufficiently long range of frequencies.}.
This is just an estimate of the variance assuming that the mean coincides with the optimal frequency.

In our analysis we did not assume constant noise variance, but took the seasonal variances 
(after removal of the rotational signal) as the estimate for the noise variances of data 
points belonging to the corresponding seasons. 
What favoured us using this kind of a noise model was the evidence of highly variable 
scatter of data in different observing seasons.
More precisely if 
$\sigma^2_i=\sigma^2_i(t_{i_1}, t_{i_2})$ is the empirical variance of the $i$-th observing 
season starting at $t_{i_1}$ and ending at $t_{i_2}$ then the noise variance $\sigma^2_{\rm n}(t_j)$ of a data point at 
time $t_j$ satisfying $t_{i_1} \le t_j \le t_{i_2}$ is set to $\sigma^2_{\rm n}(t_j) = 
\sigma^2_i$. For the seasons which had less than ten data points we used the variance of 
the full sample. The given approach still leads to a slight overestimation of the noise variance 
as seasonal segments also contain some part of the long term variations, but we 
consider this fact negligible to the results.

\begin{figure}
	\includegraphics[width=0.5\textwidth, trim={0 0.5cm 0 1cm}, clip]{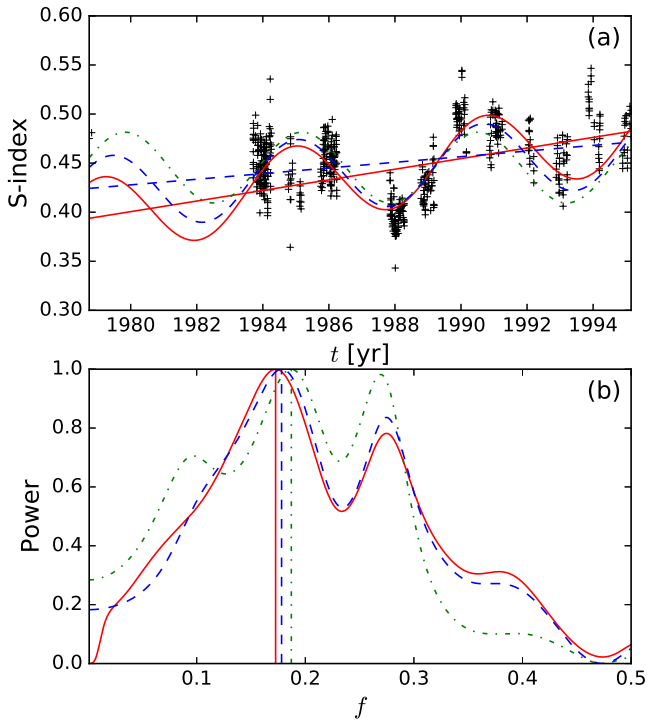}
	\caption{Comparison of the results for the star HD37394 
	using BGLST model and GLS models with and without detrending. (a) Data (black crosses), 
	BGLST model (red continuous curve), GLS model fitted to detrended 
	data with trend added back (blue dashed curve), GLS model fitted to the original data (green dash-dotted curve),
	trend component of the BGLST model (red line) and empirical trend (blue dashed line).
	(b) Spectra of the models with the same colours. 
	Vertical lines mark the locations of the corresponding maxima.}
	\label{fig_model_comp_hd37394}
\end{figure}

In Fig.~\ref{fig_model_comp_hd37394} we show the difference between the period estimates 
for the star HD37394 using the BGLST model introduced above and 
models with harmonic component only (GLS), with and without de-trending.
More examples with synthetic data can be found in 
Paper I.
It is important to note that the linear trend fitted directly to the data significantly 
differs from the trend component estimated from the BGLST method and it is clear that
all three period estimates slightly differ. 
For the reasons explained in Paper I, we assume, that the estimate from the BGLST 
method is the most reliable of the three.

\subsection{GP models}\label{GP_models}
We now turn from fully harmonic models to more complex ones, carried by the idea that due to nonlinearities 
magnetic cycles of active stars may neither be fully harmonic nor even periodic. 
One way of effectively describing any kind of time series is to use a GP model with a suitable covariance function. 

A GP is a collection of random variables which have a joint Gaussian distribution and it is
fully specified by its mean $m(\vect{x})$ and covariance $k(\vect{x}, \vect{x}^\prime)$ functions \citep{RasmussenWilliams2006}. 
GP is written as
\begin{equation}
g(\vect{x}) \sim \mathcal{GP}(m(\vect{x}), k(\vect{x}, \vect{x}^\prime)),
\end{equation}
where
\begin{equation}
\begin{aligned}
m(\vect{x})=&\mathbb{E}[g(\vect{x})]),\\
k(\vect{x}, \vect{x}^\prime)=&\mathbb{E}[(g(\vect{x}) - m(\vect{x}))(g(\vect{x}^\prime) - m(\vect{x}^\prime))].
\end{aligned}
\end{equation}
In our case the input vector $\vect{x}$ is one dimensional time $t$ and the value of S-index $y_i$ at time $t_i$ corresponds 
to the noisy observation of the function value $g(t_i)$, that is, $y_i = g(x_i) + \epsilon(t_i)$. 
We also take the mean function to be constant, that is, $m(t) = \mu$.
If we denote the vector of
observations as $\vect{y}=[y_1, y_2, \dots, y_N]\tran$, then 
\begin{equation}
	\vect{y} \sim \mathcal{N}(\mu\vect{1}, \vect{K}),
\end{equation}
where 
$\vect{1} \in \mathbb{R}^N$ is the vector of all ones and $\vect{K} \in \mathbb{R}^{N\times N}$ is the covariance matrix, whose elements
are $K_{ij} = k(t_i,t_j)+\delta_{ij}\sigma^2_{\rm n}(t_i)$. Here $\delta_{ij}$ is the Kronecker delta and $\sigma^2_{\rm n}(t)$ is 
the noise variance.

In the context of stellar rotation period estimation and variable star light curve fitting, 
GPs have recently been used more extensively \citep[see e.g.][]{Wang2012, McAllister2017, Grunblatt2015, Angus2017, Littlefair2017}.

Now we turn to the question of selecting the covariance function $k(t,t')$. The next logical step from 
fully harmonic models towards more complex ones is to drop the assumption of strict harmonicity but to allow other types of waveforms. 
In this case the GP model would be periodic with $k(t,t^\prime)=
\sigma_1^2 \exp\left(-2\sin^2(\pi f (t-t^\prime))/\ell^2\right)$, where
$\sigma_1^2$, $\ell$ and $f=1/P$ are the variance, time-scale and frequency.
However, due to the reasons discussed in 
Sect.~\ref{methods} we further add to the model the linear trend component. 
The full covariance function is then given by
\begin{equation}\label{periodic_cov}
\begin{aligned}
k(t,t^\prime)=
\sigma_1^2 \exp\left(-\frac{2\sin^2\big(\pi f (t-t^\prime)\big)}{\ell^2}\right)
+\sigma_2^2tt^\prime,
\end{aligned}
\end{equation}
where 
$\sigma_1^2$, $\ell$ and $f=1/P$ are the variance, time-scale and frequency of the periodic component, $\sigma_2^2$ is the 
variance of the trend component. The time-scale $\ell$ in the equation defines how much the form of the signal 
can deviate from the sine wave. The longer $\ell$ the more harmonic the process is and vice versa.

Another generalization towards nonharmonic models would be to assume that the process is only locally harmonic, or in other words 
the covariance function has the form of a harmonic with some damping term. One of the options for such a model would be
to use a GP with covariance function given by
\begin{equation}\label{quasiperiodic_cov}
\begin{aligned}
k(t,t^\prime)=
\sigma_1^2 \exp\left(-\frac{(t-t^\prime)^2}{2\ell^2}\right)\cos\big(2\pi (t-t^\prime) f\big)
+\sigma_2^2tt^\prime,
\end{aligned}
\end{equation}
where 
$\sigma_1^2$, $\ell$ and $f=1/P$ are the variance, 
time-scale and frequency of the quasi-periodic component and the meanings of the 
other hyperparameters are the same as in Eq.~(\ref{periodic_cov}). 
The first term in the equation is a truncated case of a more general quasi-periodic covariance function 
\footnote{This is usually 
written as the product of squared-exponential and periodic covariance functions $k(t,t^\prime)=\sigma^2 
\exp\left(-\frac{(t-t^\prime)^2}{2\ell_1^2}\right)\exp\left(-\frac{2\sin^2(\pi f (t-t^\prime))}{\ell_2^2}\right)$,
where the second exponential function can be expanded into an infinite series of cosines.
}. This 
particular simplification was made for the reason to avoid giving too much freedom to the model by reducing the number of 
parameters by one. 
The time-scale $\ell$ in the equation defines how coherent the process is. Again, the longer $\ell$ the 
more harmonic the process is and vice versa.
The main difference between the two GP models is that the first one describes a strictly periodic 
process, but possibly having a non-sinusoidal shape, while the second one describes a locally 
harmonic process. However, in the latter case, if $\ell$ is shorter than $f$
the process is dominated by red (locally correlated) noise having no strict period at all.

Next thing we need to do is to select suitable prior distributions for the hyperparameters $\sigma_1, \sigma_2, f, \ell$. This can generally be a difficult task, 
especially when there is only a small number of data points available. 
Typical of period search problems is that the likelihood of the data as function of $f$ (but not only) 
tends to be highly multimodal so that the hyperparameter optimization task becomes extremely difficult. To overcome this problem 
and reduce the computational complexity, different approximations were proposed in \citet{Wang2012}.
An alternative possibility, which we will use here, is to narrow down the hyperparameter search space by 
considering informative priors. 
In \citet{Angus2017} the authors set the GP prior as a multimodal distribution with candidate periods 
chosen by calculating the autocorrelation function of the light curve. In our case, however the 
spectra in the regions of interest are relatively smooth, because we are looking for the very long cycle 
periods. 
Thus, it suffices to assign significant probability mass to the region of 
frequency from 0 to 0.5 yrs$^{-1}$ (see Eq.~(\ref{priors})).

Yet another difficulty with applying the quasi-periodic model comes from the fact that the lengths of the 
datasets compared to the cycle periods are relatively short. Our tests showed
that the model tends to prefer solutions with small $\ell$ where the squared-exponential factor in the first term of 
Eq.~(\ref{quasiperiodic_cov}) dominates over the harmonic factor. On the contrary, we would be
interested in finding only solutions where the harmonic factor is dominating as we are 
assuming the cyclic nature of the time series after all. 
Therefore, we used total variance of the data instead of seasonal variances as the noise variance 
in the model and set a prior for $\ell$ satisfying $1/\ell \sim {\rm HalfNormal}(0, 
f/3)$. These two measures regularise the model towards smoother solutions with longer length scales.
After the optimal hyperparameters $\ell_{\rm opt}$ and $P_{\rm opt}$ are found we further reject all the 
models where $\ell_{\rm opt} < P_{\rm opt}$.
In the case of periodic GP model we imposed a less restrictive prior on $\ell$, namely $1/\ell \sim {\rm HalfNormal}(0, 
\frac{1}{6})$. This selection was made to suppress time-scales that are much shorter than 2 yrs ($3\sigma$ of the Gaussian being 
set equal to 0.5 yrs$^{-1}$).

The priors for the remaining hyperparameters are
\begin{equation}\label{priors}
\begin{aligned}
    \sigma_1^2 &\sim {\rm HalfNormal}(0, \sigma^2_{\rm m}),\\
    f &\sim {\rm HalfNormal}(0, 0.5/3),\\
    \sigma_2^2 &\sim {\rm HalfNormal}(0, \sigma^2_y/\Delta T^2),\\
    \mu &\sim \mathcal{N}(\mu_y, \sigma_y),
\end{aligned}
\end{equation}
where $\sigma^2_{\rm m}$ is the variance of the seasonal means, 
$\sigma^2_y$ is the total variance of the data, $\Delta T$ is the time span 
of the observations, 
$\mu_y = 1/N\sum_{i=1}^{N}y_i$.
The selection of priors for $f$ and $\ell$ was discussed above, but for the other hyperparameters we selected the given informative 
priors to concentrate the probability density at the places where it intuitively makes the most sense. For strictly positive 
parameters we used the positive half of the Gaussians.
The distribution of the signal variance $\sigma_1^2$ was scaled with the variance of the seasonal means $\sigma^2_m$ as we assume 
the cyclic signal to be visible primarily in the inter-seasonal variation. The prior for the trend variance $\sigma_2^2$ we scaled 
according to the rough empirical estimate of the slope the data. For the mean $\mu$ we set the expected value to the empirical mean of the 
S-index values and its variance equal to the empirical variance of the data.

For modelling and parameter inference we used the statistical library Stan\footnote{\url{http://mc-stan.org/}}. Due to limited 
computational resources we drew a relatively low number of 12800 samples 
from the joint posterior distribution of the 
parameters, from which first 6400 were 
disregarded. Tracking the number of effective samples returned by the library we concluded that this 
was, however, sufficient for our purposes.
The error estimate for the period we calculated as the standard deviation of 
the Gaussian fit to the mode of the empirical posterior distribution obtained from sampling.
We ran the sampler several times to minimize the possibility that the hyperparameter estimates would correspond to 
the local instead of the global optimum. This can happen as for multimodal distributions it is not guaranteed 
that the sampler explores the neighbourhood of all the modes.

To estimate the significance of the cycles periods retrieved by the 
GP models 
one should use red noise models, such as the GP with squared
exponential covariance function as a null hypothesis. Then one can show whether 
the given harmonic, periodic or quasi-periodic model truly holds or the data is drawn 
by pure chance from the red noise process. 
However, due to the high observational noise level, the extreme shortness and potential multicyclic
nature of the data, this kind of model selection leads to very low number of detected cycles.
Therefore we must conclude that one cannot reliably prove that the
repeating patterns in the data truly correspond to cyclic behaviour no matter
which type of waveform (harmonic, periodic, quasi-periodic) is assumed.
This also holds for the solar dataset, manifesting that data of similar quality spanning 
over equally long time span w.r.t. the cycle count also yields insignificant cycle detection,
while over longer time spans the cyclicity of its behaviour is well established.
As stars, in analogy to the Sun, are highly chaotic non-linear oscillators, we expect that
the same is true for them, that is, the significance of the cyclicity over correlated noise can be
established only with extended data sets.

As such extended data sets are not available, we estimate the significance of the retrieved cycles
from the GP models using 
a similar approach that is commonly accepted in harmonic period estimation.
We take a GP model 
with a bilinear kernel given by the last term in Eq.~(\ref{quasiperiodic_cov}) using leave one out crossvalidation (LOO-CV) on seasonal 
chunks. 
However, in the quasiperiodic case we do report whether the cycle is significant also w.r.t. to the red noise model.
LOO-CV was chosen as the Bayes factor (and therefore also BIC) is not recommended for nonparametric models due to being 
sensitive to prior definitions \citep[Chapter~5.6]{Vehtari2012}.

\subsubsection{Measure of nonharmonicity}
For the quasiperiodic GP model the period is not anymore constant.
To quantify the measure of nonharmonicity of the cycle we calculated the
so called period spread, which we will denote by $\sigma_P$.
This quantity reflects 
how much the period can on average deviate from the mean value, given that the 
time-scale $\ell$ is known exactly. 
In frequency domain this quantity is defined as the square root of 
$\sigma^2_{f}=\int_{0}^{\infty}(f-\overline{f})^2P(f)df$, where 
$\overline{f}=\int_{0}^{\infty}fP(f)df$ and $P(f)$ is the normalised power spectrum of the process \citep[see e.g.][]{Cohen1995}. 
As according to Wiener--Khinchin theorem the power spectrum of the stationary process 
is the Fourier transform of its covariance function,
for our model $\sigma_f$ is inversely proportional to $\ell$.

\subsection{Tests with quasi-periodic GP}\label{GP_QP_tests}
From now on we use shorthand H for harmonic, P for periodic GP and QP for quasi-periodic GP models.
To check the performance of the proposed QP model we first made some experiments with synthetic data.
For that reason we drew realizations from the GP with cosine covariance function with 
squared-exponential damping term added with a linear 
trend as given by Eq.~(\ref{quasiperiodic_cov}).
We varied the S/N ratio in the range from 0.25 to 4, cycle period from 2 yrs up to the 2/3 of the duration of the dataset and 
time-scale $\ell$ from half of the cycle period up to four cycle periods. We used a sampling similar to the sampling of the real data.
For each dataset we first estimated the cycle period using the harmonic model,
then we downsampled the data in a way as described in Sect.~\ref{data} and fitted QP model. 

The first observation we made from the experiments was that $\ell$ could not be very reliably estimated,
especially when the true value was longer than the total duration of the dataset. 
For shorter time-scales at least the order of magnitude of the estimate was adequate. It turned out, however, that this was sufficient 
for the QP model to yield more precise period estimates compared to the ones from H model.

As a diagnostic we choose the relative error of the frequency estimate $\Delta = |f_{\rm est} - f_{\rm true}|/f_{\rm true}$, where 
$f_{\rm est}$ is the estimated cycle frequency and $f_{\rm true}$ is the true cycle frequency. For each value of $\ell$ less than a 
chosen limit we calculated the relative errors for both QP and H estimates $\Delta_{\rm GP}$ and $\Delta_{\rm H}$ and 
their corresponding mean values over the set of experiments denoted by $\overline{\Delta}$.

\begin{figure}
	\includegraphics[width=0.5\textwidth, trim={0 1cm 0 2cm}, clip]{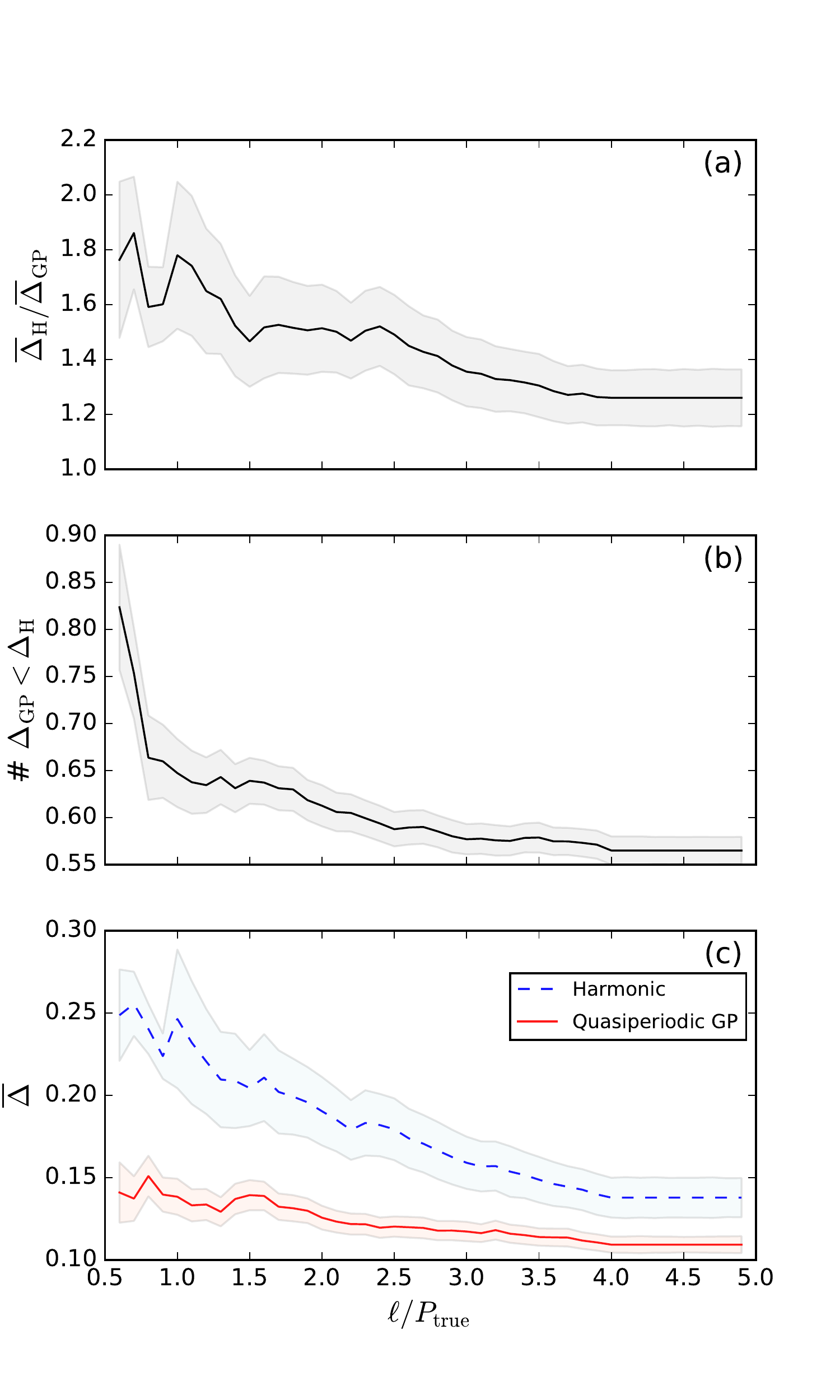}
	\caption{Diagnostics of the experiments comparing QP and H models. (a) The ratio of average relative 
		errors, (b) the fraction of experiments where GP outperformed 
		harmonic model, (c) relative errors of QP (red continuous curve) and H (blue dashed curve) models.}\label{fig_exp}
\end{figure}

From Fig.~\ref{fig_exp}(a) we see that, on average, the estimate from QP
is more accurate than the one from H, but 
this difference decreases towards the longer time-scales. 
This is easy to understand, as the signals with longer time-scales are more harmonic.
Similarly from Fig.~\ref{fig_exp}(b) we see that the fraction of experiments where the period estimates from QP were 
more accurate than the estimates from H is larger than 50\% for time-scales in the 
range from 0.5 to 5. 
In Fig.~\ref{fig_exp}(c) we show the average relative 
errors of period separately for QP (red continuous curve) and H (blue dashed curve) models. Again we see that the QP 
outperforms H, while the difference is more pronounced for short time-scales than 
for very long ones. 
The reason why relative errors are large even for the longer values of $\ell$ 
(always above 10\%) could be explained by 
the biased estimates in the experiments with low S/N ratio and/or bad sampling.

From the experiments we conclude that in the case of not strictly harmonic processes and regardless of the
datasets being extremely short, one can use QP model to obtain more 
accurate period estimate compared to the estimates from the fully harmonic model.
This is regardless of the fact that for the QP model downsampled datasets were used.
This observation motivated us to actually use the QP model on real data. 
We did not repeat the tests with the 
periodic covariance function, believing that the results 
differ less compared to the harmonic model, especially when the true process is quasi-periodic.

We note that in these experiments, both the form of the true covariance function and the one used in the QP model coincided. In 
practice, however, the form of the true covariance function is usually unknown and can only be guessed based on some physical 
arguments. 
Drawn from the arguments given in the beginning of Sect.~\ref{methods}, we have a strong belief that 
a more natural choice for modelling time series of stellar activity would be with quasi-periodic
covariance function rather than with harmonic or periodic ones. Which exact form of the quasi-periodic covariance function to 
use depends, however, on the situation. 
In our case, as we briefly commented above, we decided to keep the covariance function as simple as possible, 
building it up from the widely used squared-exponential damping and cosine terms. 
We did not consider other forms of the damping term, 
but one of the well known alternatives would be given by $\exp(-|t-t^\prime|/\ell)$. It has the
desirable property of  
the computational cost of fitting the model scaling linearly with 
the number of data points \citep{Foreman-Mackey2017}.
We did not consider using this approach, however, because we needed to incorporate the linear trend into the model as an additional 
component. 

In the experiments with synthetic data we knew the true cycle frequencies which allowed us to
calculate the actual errors.
In Sect.~\ref{results} we give error estimates for the real datasets,
which are calculated as the standard deviations of the highest modes 
of the posterior distributions. One could expect these quantities to give 
information on the possible values of the true errors, however, due to
additional uncertainties (e.g. whether the process can be assumed to be Gaussian
and the form of the covariance function is correct)
in practice the errors are most likely underestimated.

\section{Results}\label{results}
When analysing the real datasets we rejected all the cycle lengths (regardless of the significance level) 
below two years or longer than 2/3 of the total time span of the dataset.
The reason for the lower limit comes from the arguments related to the seasonal sampling patterns in the data 
as well as from the need to avoid falling into the domain of rotational periods of the stars. The longest 
known rotational period in our dataset is around 163 days, but for some of the stars the estimate is missing.
Setting the lower limit prohibits us from finding for example the 120 d cycle of $\tau$ Boo reported by
\citep{Mittag2017, Jeffers2018} (clearly distinct from its $P_{\rm rot}=3.07$ d rotation period),
but it is a necessary constraint for performing a uniform cycle search for
all the stars in our sample.
The upper limit was fixed for the sake of not making too light conclusions on the existence of long cycles. 
Obviously by using the given period search range we also fail to detect very long cycles.
However, as even in the best cases we see only a couple of full cycles, 
all the given estimates, especially those with lower significance level, should be taken with some caution. 
Only future observations can make the estimates of these cycles more accurate. 

\subsection{Summary of the cycles}
\begin{figure}
	\includegraphics[width=0.5\textwidth, trim={0 1cm 0 2cm}, clip]{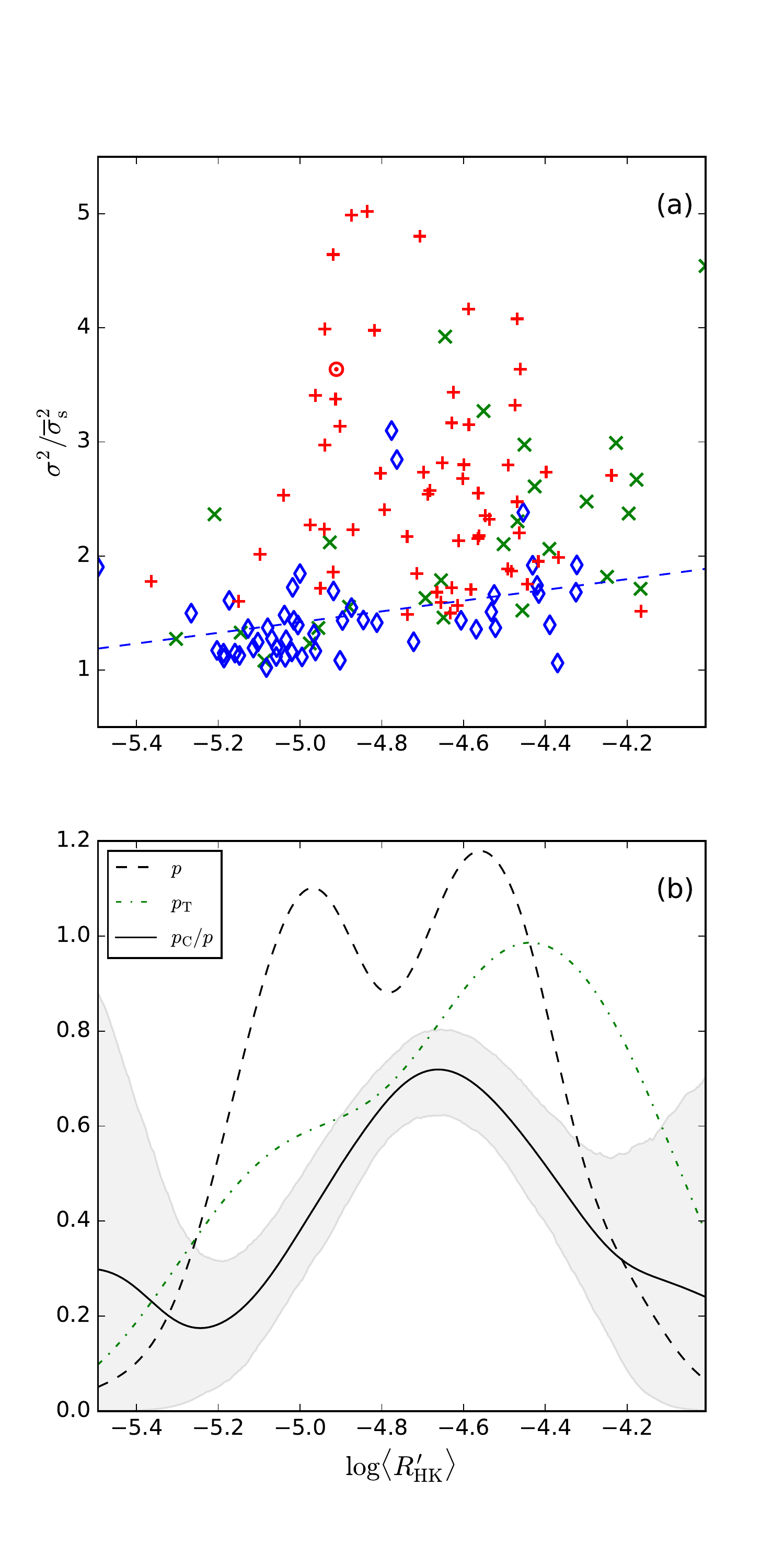}
	\caption{(a) Ratio of total variance to average seasonal variance vs. 
		$\log$\RHK. NC stars (blue diamonds), T stars
		(green crosses), C stars (red pluses). Blue dashed line represents the linear fit to the data of NC stars. 
		(b) $p$ is the density (i.e. number of stars per unit $\log$\RHK) of all stars, 
		$p_{\rm C}$ is the density of the stars with cycles
		and $p_{\rm T}$ is the density of the stars having a linear trend.
		The solid black curve represents the percentage of C stars as 
		function of $\log$\RHK.
		}
	\label{cyc_vs_act}
\end{figure}

In summary, we find 61 stars with cycles (hereafter class C),
the cycle periods, with their error estimates and significances
for all the three methods used here
being gathered into Tables~\ref{cycles_1} and \ref{cycles_2}.
In addition, we find 26 stars with trends (hereafter class T), that may represent cycles
longer than can be reliably detected from the current data.
The sample contains 49 noncyclic stars (hereafter class NC) that do not show any trend either.
Stars in classes T and NC are not listed in any table, but they are included in the plots
in Fig.~\ref{cyc_vs_act}.

A measure of
total variability of the stars is shown in Fig.~\ref{cyc_vs_act}(a). 
We define total variability as the ratio of
total variance $\sigma^2$ and average seasonal variance $\overline{\sigma}^2_{\rm s}$.
The idea of this plot is to indicate how much irregular inter-seasonal variability occurs in the stars;
constant noise level would be manifested by this quantity being close to one.
The less active stars belonging to class NC all cluster close to the value of one, or slightly above, and only weak scatter is observed.
The more active stars of class NC exhibit larger values (between 1--2) of total variability (even up to 3), and
show much larger scatter.
A linear fit to all stars in class NC (shown with a dashed line) clearly demonstrates this behaviour.
We interpret this as the
active stars having significantly increased tendency to show irregular variability, that cannot be detected nor characterised
with the available statistical toolboxes from the given data.

\begin{landscape}
	\begin{table}\caption{Stars with cycles agreeing with either harmonic or GP models. The meanings of the columns read as follows: ID is the HD 
			identifier of the star, $\Delta T$ is the time duration of the observation in years, ST/L is the spectral type with luminosity class, MS 
			indicates whether the star belongs to main-sequence, 
			$P_{\rm rot}$ is the rotational period of the star, $\log$\RHK is the logarithm of average chromospheric activity index,
			$P_{\rm H}$, $P_{\rm P}$, $P_{\rm QP}$ are correspondingly the cycle periods from H, 
			P and QP models, $\Delta P$ is the standard deviation of the period, $\Delta$BIC and 
			$\Delta$LOO-CV show the 
			significance of the model. 
			In the QP case the cycle period is shown with boldface font if the model was significant w.r.t. the red noise model.
				$\sigma_P$ indicates the spread of the period due to nonharmonicity.
			In the last column we have shown the results from Cyc95, where the letter in 
			brackets refers to the grade of the cycle: E - excellent, G - good, F - fair and P - poor. The boldface font of the HD 
			identifier indicates the extended dataset.
			Throughout the table `--' means that no cycle was found and `\dots' indicates 
			that the star was not included in the given study, or the value in the given column is not known. 
			Question marks refer to unreliable estimates.}
		\begin{tabular}{l|l|l|l|l|l|l|l|l|l|l}
			ID & $\Delta T$ & ST/L & MS & $P_{\rm rot}$ & ${\rm log} \langle R^\prime_{\rm HK}\rangle$ &$P_{\rm H} \pm \Delta P$ 
			($\Delta$BIC) & 
			\begin{tabular}{@{}c@{}} $P_{\rm P} \pm \Delta P$ \\ ($\Delta$LOO-CV) \end{tabular}
			& \begin{tabular}{@{}c@{}} $P_{\rm QP} \pm \Delta P$ \\ ($\Delta$LOO-CV) \end{tabular} & $\sigma_P$ & Cyc95 \\ \hline
			HD100180 & 27.3 & F V & \cmark & 13.51 $\pm$ 0.24 & -4.918 & -- & -- & -- & -- &  3.56 (F) 12.9 (F) \\ 
			HD10072 & 11.4 & G III & \xmark & 122.1 $\pm$ 9.0 & -4.562 &  6.04 $\pm$ 0.2 (9.8) & -- &  7.25 $\pm$ 0.65 (28.4) & 0.1 & $\dots$ \\ 
			HD101501 & 27.3 & G V & \cmark & 15.95 $\pm$ 0.2 & -4.587 & -- &  12.75 $\pm$ 0.07 (31.0) & -- & -- & -- \\ 
			{\bf HD103095} & 33.3 & K V & \cmark & 34.03 $\pm$ 0.68 & -4.939 &  7.13 $\pm$ 0.03 (86.4) &  21.33 $\pm$ 3.68 (46.0) &  7.21 $\pm$ 0.17 (59.4) & 0.0 &  7.3 (E) \\ 
			{\bf HD10476} & 35.1 & K V & \cmark & 35.6 $\pm$ 0.75 & -4.962 &  10.6 $\pm$ 0.08 (110.4) & -- &  10.14 $\pm$ 0.15 (97.4) & 0.5 &  9.6 (E) \\ 
			HD10780 & 16.2 & K V & \cmark & 22.14 $\pm$ 0.55 & -4.707 &  7.53 $\pm$ 0.16 (98.7) & -- &  10.22 $\pm$ 0.61 (81.5) & 0.1 &  >7 \\ 
			HD111456 & 16.1 & F V & \cmark & $\dots$ & -4.37 & -- & -- & -- & -- &  7? \\ 
			{\bf HD114710} & 35.2 & F V & \cmark & 11.99 $\pm$ 0.1 & -4.738 &  16.56 $\pm$ 0.22 (8.6) &  16.34 $\pm$ 2.31 (15.8) &  14.12 $\pm$ 1.79 (9.0) & 0.3 &  16.6 (G) 9.6 (F) \\ 
			HD115043 & 17.1 & G V & \cmark & 5.51 $\pm$ 0.1 & -4.444 & -- & -- & {\bf 2.94} $\pm$ 0.11 (11.5) & 0.2 & -- \\ 
			{\bf HD115383} & 35.2 & G V & \cmark & 3.551 $\pm$ 0.046 & -4.464 & -- & -- &  16.95 $\pm$ 0.38 (15.3) & 0.0 & -- \\ 
			{\bf HD115404} & 33.3 & K V & \cmark & 18.03 $\pm$ 0.26 & -4.502 & -- & -- & -- & -- &  12.4 (G) \\ 
			{\bf HD120136} & 34.3 & F IV & \cmark & 3.07 $\pm$ 0.062 & -4.722 & -- & -- & -- & -- &  11.6 (P) \\ 
			HD124897 & 11.4 & K III    - & \xmark & 261.7 $\pm$ 7.0 & -5.15 & -- &  5.38 $\pm$ 0.11 (10.5) & -- & -- & $\dots$ \\ 
			HD126053 & 29.3 & G V & \cmark & 25.7 $\pm$ 1.7 & -4.966 & -- & -- & -- & -- &  22? \\ 
			{\bf HD131156A} & 34.0 & G V & \cmark & 6.0 $\pm$ 0.036 & -4.368 & -- & -- & {\bf 17.2} $\pm$ 2.13 (6.6) & 1.5 & -- \\ 
			{\bf HD146233} & 20.1 & G V & \cmark & 22.62 $\pm$ 0.57 & -4.95 &  11.2 $\pm$ 0.4 (9.5) &  11.42 $\pm$ 0.63 (14.7) & {\bf 11.6} $\pm$ 0.25 (14.7) & 0.1 & $\dots$ \\ 
			{\bf HD149661} & 34.1 & K V & \cmark & 20.76 $\pm$ 0.28 & -4.625 & -- &  12.31 $\pm$ 0.06 (19.0) & -- & -- &  17.4 (G) 4 (G) \\ 
			{\bf HD152391} & 35.0 & G V & \cmark & 10.62 $\pm$ 0.13 & -4.469 & \begin{tabular}[t]{@{}l@{}} 9.03 $\pm$ 0.05 (8.7)\\  13.73 $\pm$ 0.18 (7.6)\end{tabular} & -- &  9.85 $\pm$ 0.79 (48.3) & 1.1 &  10.9 (E) \\ 
			HD154417 & 29.3 & F V & \cmark & 7.812 $\pm$ 0.062 & -4.537 & -- & -- & {\bf 9.52} $\pm$ 0.21 (22.3) & 1.3 &  7.4 (F) \\ 
			HD155885 & 28.0 & K V & \cmark & 18.87 $\pm$ 0.42 & -4.564 & -- &  17.53 $\pm$ 1.99 (10.6) &  5.73 $\pm$ 0.15 (13.5) & 0.6 &  5.7 (P) \\ 
			HD155886 & 28.0 & K V & \cmark & 20.58 $\pm$ 0.5 & -4.599 & \begin{tabular}[t]{@{}l@{}} 10.44 $\pm$ 0.21 (18.6)\\  5.0 $\pm$ 0.04 (16.9)\end{tabular} & -- &  13.71 $\pm$ 1.56 (23.7) & 0.1 & -- \\ 
			HD156026 & 28.9 & K V & \cmark & 16.69 $\pm$ 0.66 & -4.612 &  17.89 $\pm$ 0.51 (10.8) &  17.72 $\pm$ 0.74 (28.1) & -- & -- &  21 (G) \\ 
			{\bf HD157856} & 35.0 & F V & \xmark & 13.3 $\pm$ 0.34 & -4.629 & -- & -- &  11.73 $\pm$ 0.2 (14.0) & 0.7 &  15.9 (F) \\ 
			{\bf HD160346} & 35.1 & K V & \cmark & 32.0 $\pm$ 1.0 & -4.818 &  7.21 $\pm$ 0.03 (265.8) &  7.18 $\pm$ 0.01 (290.8) &  7.14 $\pm$ 0.19 (249.3) & 0.0 &  7 (E) \\ 
			{\bf HD161239} & 35.1 & G IV & \xmark & 23.33 $\pm$ 0.5 & -5.266 & -- & -- & -- & -- &  5.7 (F) 11.8 (P) \\ 
			HD16160 & 27.3 & K V & \cmark & 48.58 $\pm$ 0.8 & -4.902 &  12.45 $\pm$ 0.14 (154.2) &  12.62 $\pm$ 1.4 (109.4) &  12.62 $\pm$ 0.2 (107.3) & 0.0 &  13.2 (E) \\ 
			HD165341A & 24.8 & K V & \cmark & 19.33 $\pm$ 0.31 & -4.565 &  5.19 $\pm$ 0.03 (30.9) & -- &  5.86 $\pm$ 0.28 (100.2) & 0.4 &  5.1 (F) \\ 
			HD165341B & 23.2 & K V & \xmark & $\dots$ & $\dots$ & -- & -- & {\bf 15.23} $\pm$ 0.63 (12.1) & 0.1 & -- \\ 
			HD166620 & 28.8 & K V & \cmark & 42.1 $\pm$ 1.3 & -4.975 &  16.16 $\pm$ 0.3 (58.5) &  16.89 $\pm$ 0.19 (54.8) &  16.62 $\pm$ 0.18 (58.9) & 0.1 &  15.8 (E) \\ 
			HD16673 & 28.5 & F V & \cmark & 5.976 $\pm$ 0.09 & -4.632 & -- &  5.6 $\pm$ 0.68 (11.5) & -- & -- & -- \\ 
			HD176051 & 17.1 & F V & \cmark & 15.37 $\pm$ 0.39 & -4.874 & -- & -- & -- & -- &  10? \\ 
			HD182101 & 28.4 & F V & \cmark & 4.774 $\pm$ 0.058 & -4.569 & -- & -- & -- & -- &  5.1 (P) \\ 
			{\bf HD18256} & 35.1 & F V & \cmark & 3.648 $\pm$ 0.027 & -4.714 & -- &  6.64 $\pm$ 0.06 (13.1) &  6.18 $\pm$ 0.21 (12.0) & 0.3 &  6.8 (F) \\ 
			{\bf HD1835} & 35.2 & G V & \cmark & 7.676 $\pm$ 0.053 & -4.454 & -- & -- & -- & -- &  9.1 (F) \\ 
			HD185144 & 18.2 & G V & \cmark & 27.7 $\pm$ 0.77 & -4.836 &  6.66 $\pm$ 0.05 (131.7) &  6.54 $\pm$ 0.02 (98.1) &  6.93 $\pm$ 0.56 (82.4) & 0.3 &  7? \\ 
			HD187691 & 28.4 & F V & \cmark & 10.38 $\pm$ 0.16 & -5.018 & -- & -- & -- & -- &  5.4 (F) \\ 
			HD188512 & 16.5 & G IV & \xmark & $\dots$ & -5.173 & -- & -- & -- & -- &  4.1 (P) \\ 
		\end{tabular}\label{cycles_1}
	\end{table}
\end{landscape}

\begin{landscape}
	\begin{table}\caption{Stars with cycles (cont.)}
		\begin{tabular}{l|l|l|l|l|l|l|l|l|l|l}
			ID & $\Delta T$ & ST/L & MS & $P_{\rm rot}$ & ${\rm log} \langle R^\prime_{\rm HK}\rangle$ &$P_{\rm H} \pm \Delta P$ 
			($\Delta$BIC) & 
			\begin{tabular}{@{}c@{}} $P_{\rm P} \pm \Delta P$ \\ ($\Delta$LOO-CV) \end{tabular}
			& \begin{tabular}{@{}c@{}} $P_{\rm QP} \pm \Delta P$ \\ ($\Delta$LOO-CV) \end{tabular} & $\sigma_P$ & Cyc95 \\ \hline
			{\bf HD190007} & 34.1 & K V & \cmark & 27.68 $\pm$ 0.36 & -4.652 & -- &  16.58 $\pm$ 3.27 (17.9) &  16.36 $\pm$ 0.36 (7.0) & 0.1 &  13.7 (F) 5.3 (P) \\ 
			HD190406 & 28.5 & G V & \cmark & 14.09 $\pm$ 0.21 & -4.793 & -- &  7.82 $\pm$ 0.79 (14.2) &  16.99 $\pm$ 3.79 (15.9) & 0.3 &  2.6 (F) 16.9 (G) \\ 
			HD194012 & 28.5 & F V & \cmark & 6.73 $\pm$ 0.11 & -4.665 & -- & -- &  3.43 $\pm$ 0.1 (10.7) & 0.4 &  16.7 (P) 5.4 (F) \\ 
			{\bf HD201091} & 34.3 & K V & \cmark & 35.54 $\pm$ 0.47 & -4.588 & \begin{tabular}[t]{@{}l@{}} 7.16 $\pm$ 0.02 (118.3)\\  21.09 $\pm$ 0.31 (19.2)\end{tabular} & -- &  8.59 $\pm$ 0.63 (62.0) & 0.1 &  7.3 (E) \\ 
			{\bf HD201092} & 34.3 & K V & \cmark & 34.55 $\pm$ 0.57 & -4.803 &  11.65 $\pm$ 0.1 (22.3) & -- & -- & -- &  11.7 (G) \\ 
			HD20630 & 27.3 & G V & \cmark & 8.966 $\pm$ 0.094 & -4.468 & -- & -- & -- & -- &  5.6 (F) \\ 
			HD206860 & 28.6 & G V & \cmark & 4.851 $\pm$ 0.048 & -4.416 & -- & -- & -- & -- &  6.2 (P) \\ 
			HD218658 & 11.3 & G III & \xmark & 4.607 $\pm$ 0.048 & -4.398 & -- &  6.83 $\pm$ 0.05 (10.5) & -- & -- & $\dots$ \\ 
			HD219834A & 27.3 & G IV & \xmark & 43.4 $\pm$ 1.9 & -5.098 &  16.29 $\pm$ 0.35 (9.2) & -- & -- & -- &  21 (G) \\ 
			HD219834B & 27.3 & K IV? & \xmark & 34.78 $\pm$ 0.86 & -4.919 &  9.32 $\pm$ 0.08 (25.2) &  9.76 $\pm$ 0.83 (36.8) &  9.56 $\pm$ 0.12 (33.4) & 0.0 &  10 (E) \\ 
			HD22049 & 27.3 & K V & \cmark & 11.09 $\pm$ 0.19 & -4.461 & -- &  17.15 $\pm$ 3.34 (10.5) & {\bf 11.0} $\pm$ 1.02 (7.1) & 0.4 & -- \\ 
			HD224930 & 28.5 & G V & \cmark & 30.19 $\pm$ 0.95 & -4.913 & -- & -- & {\bf 16.85} $\pm$ 2.0 (39.6) & 2.0 &  10.2 (P) \\ 
			HD2454 & 28.4 & F V & \cmark & 3.041 $\pm$ 0.072 & -4.737 & -- &  3.47 $\pm$ 0.01 (8.3) & -- & -- & -- \\ 
			HD26913 & 28.4 & G V & \cmark & 7.028 $\pm$ 0.077 & -4.417 & -- & -- &  10.03 $\pm$ 0.14 (9.6) & 0.8 &  7.8 (F) \\ 
			HD26923 & 28.4 & G IV & \cmark & 10.6 $\pm$ 0.25 & -4.492 & -- & -- &  7.61 $\pm$ 0.43 (22.6) & 0.7 & -- \\ 
			HD26965 & 27.3 & K V & \cmark & 38.65 $\pm$ 0.58 & -4.919 &  10.66 $\pm$ 0.11 (229.0) &  10.36 $\pm$ 0.06 (203.2) &  10.31 $\pm$ 0.26 (227.6) & 0.0 &  10.1 (E) \\ 
			HD27022 & 11.5 & G II & \xmark & $\dots$ & -4.469 &  5.65 $\pm$ 0.18 (42.6) & -- & {\bf 5.74} $\pm$ 0.18 (72.1) & 0.0 & $\dots$ \\ 
			HD29317 & 15.2 & K III & \xmark & 128.2 $\pm$ 1.5 & -4.87 & \begin{tabular}[t]{@{}l@{}} 7.95? (103.4)\\  6.75 $\pm$ 0.14 (8.5)\end{tabular} & -- & {\bf 9.93} $\pm$ 1.13 (136.0) & 0.1 & $\dots$ \\ 
			HD32147 & 27.3 & K V & \cmark & 33.7 $\pm$ 1.1 & -4.939 &  11.13 $\pm$ 0.12 (311.4) &  11.06 $\pm$ 0.05 (240.4) & {\bf 12.86} $\pm$ 0.7 (214.7) & 0.5 &  11.1 (E) \\ 
			HD3229 & 28.4 & F V & \xmark & 1.525 $\pm$ 0.064 & -4.532 & -- & -- & -- & -- &  4.9 (P) \\ 
			HD33608 & 28.1 & F V & \cmark & 3.211 $\pm$ 0.073 & -4.582 & -- & -- &  6.13 $\pm$ 0.41 (8.3) & 0.7 & -- \\ 
			HD3651 & 28.4 & K V & \cmark & 37.0 $\pm$ 1.2 & -5.04 &  16.98 $\pm$ 0.39 (41.1) & -- &  12.45 $\pm$ 0.67 (54.4) & 0.3 &  13.8 (G) \\ 
			HD37394 & 16.4 & K V & \cmark & 11.49 $\pm$ 0.22 & -4.474 &  5.83 $\pm$ 0.08 (20.2) & -- &  3.77 $\pm$ 0.16 (6.3) & 0.0 &  3.6 (P) \\ 
			HD4628 & 28.5 & K V & \cmark & 37.14 $\pm$ 0.62 & -4.874 & \begin{tabular}[t]{@{}l@{}} 8.56 $\pm$ 0.06 (212.2)\\  5.79 $\pm$ 0.05 (9.4)\end{tabular} &  8.41 $\pm$ 0.03 (151.9) &  8.55 $\pm$ 0.22 (147.7) & 0.0 &  8.37 (E) \\ 
			HD57727 & 9.5 & G III & \xmark & 87.3 $\pm$ 2.3 & -4.629 & -- &  5.75 $\pm$ 0.13 (20.2) & -- & -- & $\dots$ \\ 
			HD60522 & 10.5 & M III & \xmark & $\dots$ & -5.364 &  5.61? (21.0) &  5.48 $\pm$ 0.09 (48.8) & {\bf 4.98} $\pm$ 0.65 (25.7) & 0.0 & $\dots$ \\ 
			HD68290 & 11.4 & K III & \xmark & 158.4 $\pm$ 2.8 & -4.682 &  5.51 $\pm$ 0.12 (16.6) &  5.42 $\pm$ 0.08 (17.9) & {\bf 5.75} $\pm$ 0.2 (19.4) & 0.0 & $\dots$ \\ 
			{\bf HD75332} & 35.1 & F V & \cmark & 3.673 $\pm$ 0.099 & -4.483 & -- &  22.8 $\pm$ 0.47 (11.7) & -- & -- & -- \\ 
			HD76151 & 29.1 & G V & \cmark & 14.4 $\pm$ 0.19 & -4.698 & \begin{tabular}[t]{@{}l@{}} 15.9 $\pm$ 0.38 (17.7)\\  5.09 $\pm$ 0.04 (10.2)\end{tabular} & -- & -- & -- &  2.52 (F) \\ 
			{\bf HD76572} & 35.0 & F V & \xmark & 6.77 $\pm$ 0.16 & -4.896 & -- & -- & -- & -- &  7.1 (P) \\ 
			HD78366 & 29.1 & G IV-V & \cmark & 9.519 $\pm$ 0.079 & -4.602 &  14.63 $\pm$ 0.24 (21.3) & -- & -- & -- &  12.2 (G) 5.9 (F) \\ 
			{\bf HD81809} & 35.1 & G V & \xmark & 41.66 $\pm$ 0.8 & -4.94 &  8.11 $\pm$ 0.04 (67.5) &  8.25 $\pm$ 0.04 (60.2) &  8.16 $\pm$ 0.09 (63.6) & 0.1 &  8.17 (E) \\ 
			HD82443 & 16.3 & G V & \cmark & 5.39 $\pm$ 0.1 & -4.238 & -- &  5.58 $\pm$ 0.08 (12.2) & {\bf 2.79} $\pm$ 0.11 (10.4) & 0.0 &  2.8 (P) \\ 
			HD82635 & 15.4 & G III & \xmark & 76.6 $\pm$ 1.9 & -4.491 &  8.79 $\pm$ 0.36 (54.9) &  9.26 $\pm$ 0.18 (72.6) &  9.44 $\pm$ 0.34 (71.7) & 0.1 & $\dots$ \\ 
			{\bf HD82885} & 33.2 & G V & \cmark & 17.88 $\pm$ 0.18 & -4.687 & -- &  17.07 $\pm$ 3.02 (14.9) & {\bf 10.85} $\pm$ 0.26 (17.1) & 1.3 &  7.9 (F) 12.6 (P) \\ 
			HD85444 & 11.4 & G III & \xmark & 91.2 $\pm$ 5.1 & -4.547 & -- &  6.65 $\pm$ 0.08 (19.0) & -- & -- & $\dots$ \\ 
			HD88373 & 25.1 & F V & \xmark & 6.6 $\pm$ 0.18 & -4.615 & -- & -- &  11.68 $\pm$ 0.27 (6.4) & 0.2 & $\dots$ \\ 
			HD88737 & 29.2 & F V & \xmark & 6.85 $\pm$ 0.13 & -4.655 & -- &  11.49 $\pm$ 0.21 (8.6) & {\bf 12.25} $\pm$ 1.66 (7.3) & 0.1 &  24? \\ 
			HD89449 & 11.3 & F IV-V & \cmark & $\dots$ & -4.166 & -- &  2.76 $\pm$ 0.04 (14.5) & -- & -- & $\dots$ \\ 
			{\bf Sun} & 49.1 & G V & \cmark & 26.09 & -4.911 &  10.89 $\pm$ 0.03 (86.6) &  11.01 $\pm$ 0.03 (67.9) &  10.57 $\pm$ 0.41 (96.8) & 1.1 &  10.0 (E) \\ 
		\end{tabular}\label{cycles_2}
	\end{table}
\end{landscape}

In the light of dynamo theory, the less systematic behaviour of the active stars could simply be
interpreted as their dynamos operating in a more supercritical regime,
assuming that the field generators are increasing in magnitude as function of rotation,
therefore showing a tendency for more complex solutions, as proposed by \citet{Durney1981}.
Class C stars do not show any clear trends in Fig.~\ref{cyc_vs_act}(a),
but the overall scatter of points is higher than for NC stars indicating the fact that strong inter-seasonal
variations are always explained by cyclic, not irregular behaviour.
The largest variability is seen for the C stars that are located in the mid-range of
\RHK, where the overall density of stars is significantly decreased (see next paragraph).

We also plot the histograms of all the stars, and separately for classes C and T, over $\log$\RHK\ in 
Fig.~\ref{cyc_vs_act}(b).
We clearly see that the distribution of all the stars in this sample is
bimodal, there being a clear minimum around the activity index value $-4.8$. This `gap', which is not totally void of stars,
divides them into two populations, a distribution with active and inactive stars. The
minimum seen corresponds to the Vaughan-Preston
gap (VPG), that is the decreased abundance
of stars in a certain range of the chromospheric activity index \citep{VP80}.
The Sun with $\log$\RHK $\approx -4.911$ is located nearby 
the region of the minimum (or gap).
We note that in the larger catalogue of nearly 4,500 chromospherically active stars, 
compiled by \cite{BS18}, the VPG is, however, much less pronounced.

Comparing the relatively low probability to find either C or T stars among the inactive stars with
the distribution of all the stars, we can conclude that NC stars form the 
major part of the inactive population.
Class C shows an opposite trend: there is an increased probability to find a cyclic star 
close to the gap in the active population than
elsewhere. However, as stated before, more prominent cycles seem to appear near the `gap'.
Also, the active population shows more stars with a trend, indicative of the presence of many undetectable long cycles in this
population. In that sense, the cycles discussed later on for this population might not be truly indicative of these stars, but
their cycles need even longer time spans to be detected properly.

\subsection{Comparison with Cyc95}\label{results_comp_cyc95}

Next we turn into more detailed comparison of the cycle length estimates found in the current 
study to those of Cyc95.
We note here that the stars excluded from Tables \ref{cycles_1} and \ref{cycles_2}
have no cycles detected from neither of the studies and
the datasets, for which we have ten additional observing seasons compared to Cyc95,
are highlighted using boldface font in the
first columns of these tables. For the rest of the datasets we have only four additional seasons.

The first 
observation we make is that the cycle estimates for the stars,
which Cyc95 classified having  
`excellent' cycle grades, agree reasonably well between both studies. 
Small discrepancy can be found only for the star HD152391 where there were two cycles 
detected in our study, but neither of them matches the single estimate from Cyc95. Here we note, 
however, that the difference can be explained by the fact that for the given star 
ten years longer dataset was available to us.

For stars with `good' grades, we find many more discrepancies between the results.
The extreme case is HD115404,
for which no cycle was detected at all from our study. 
Here again our dataset was ten years longer compared to the older study.
Another significant discrepancy takes place for 
HD3651, where the cycle period from H
is longer (however, the QP estimate is shorter).
In Cyc95 the trend was removed beforehand, while in our case it is fit to the data 
simultaneously with the harmonic component. 

Next we proceed to the comparison of stars with `fair'
grade.
For them we see agreement in only one occasion, namely
for HD165341A do the estimates coincide.
Mostly there is neither a cycle detected from our methods or the estimated cycle lengths are differing. The 
reason for the former is that we used a rather strong significance level and rejected all weak estimates.
There is one exception here, namely HD18256, in which case H did not detect the cycle,
but the estimates from P and QP match well with the estimate from Cyc95.

Lastly, for none of the `poor' cycles except for 
HD37394 did we find a cycle using H, but even in this case the estimated cycle lengths are differing.
Interestingly, QP in this case detected very close value to the one from Cyc95,
however, in most of the other cases the GP estimates strongly differ from Cyc95 as well.

The number of double harmonics found from both of the studies is different. There are total of nine stars reported 
in Cyc95 with double harmonicity, but excluding the stars with poor cycle 
estimates, this number reduces to 5, which matches the number of double harmonics
found in our study. However, none of the 
stars with double periodicities match, indicating that the treatment of the trends and the inclusion of new
data have altered the picture of double periodicities very significantly. Therefore, these earlier
detections appear unreliable.
The most striking disagreement
is found for HD149661.
In the previous study there were two good cycles found, but no significant cycles from H at all in the
current study, where we had ten years longer dataset available. 
There is an estimate found from P, but this is not agreeing with neither of the estimates from Cyc95.
Furthermore there is a star HD155886 with double cyclicity detected
in the present study with no clear cycle from
Cyc95, although there was a remark made on the presence of significant variability in the data.

Another difference between the studies is that the frequencies of 
double harmonics found in Cyc95 form irrational ratios,
while in our study they tend to form ratios of integer multiples.
For six of the stars our harmonic model detected two separate periods,
and in the case of three of them (HD155886, HD201091 and HD76151),
the two cycle periods were integer multiples of each other within given uncertainty
limits. For the  former star the ratio is roughly two and for the latter two ones
roughly 3, whereas for the star HD201091 the higher harmonic is the more prominent one. 
We argue that the periods whose ratios form integer numbers cannot be considered as different
magnetic cycles, but as the harmonics of the same cycle.
Usually these harmonics appear as the weaker overtones of the harmonic with the basic frequency, 
however, due to sub-harmonic bifurcation this needs not always be the case. 

For two of the stars for which we detected irrational cycle fractions using the harmonic model, 
the period estimates are quite close to each other. Due to the poor spectral resolution,
especially for HD29317 because of very poor sampling,
we cannot state if there are actually two approximately harmonic cycles or one
less coherent one.
At least the cycle estimate from QP being between the two estimates from H for HD152391 seems to confirm the latter case

Finally, there is a
number of stars with short time series which have not been analysed by Cyc95, but for which we 
report significant cycles at least from two analysis methods. These stars include HD10072, HD27022, HD29317, 
HD60522, HD68290, HD82635 and HD146233. 

To summarise this subsection,
the biggest differences between the studies originate from two causes.
Firstly, the linear trend components retrieved by our models significantly differ from zero,
from the trends obtained directly from linear regression.
The fact that the trends differ also from star to star indicates that they are most likely not due to systematic instrumental trend.
This is the major reason for the different cycle length estimates obtained in the case of good, fair,
and poor cycle grades of Cyc95.
Secondly, the extended lengths of the datasets available to us enabled
improving of some cycle length estimates.

\subsection{Differences between the models}\label{results_comp_models}
Next we summarise the differences between
H, P and QP models.
The primary observation we make is the fact that QP detected slightly more cycles
as the other methods. We note that the model selection methods differ between the H and GP
methods;
hence the corresponding significance levels shown in Tables~\ref{cycles_1} and \ref{cycles_2} 
are not directly comparable. However, as the numbers for the strong cycles are roughly
speaking equal, we chose the same cut-off level for the GP models as for H.
We therefore find the most likely explanation to the higher number of cycles from QP to be 
the increased flexibility of the model because of the squared-exponential damping term in 
the covariance function.

We omit detailed investigation of differences between the models per each individual star, but highlight some of the
interesting cases.
First, there are plenty of stars for which P did detect cycle, but H did not and vice versa. 
The reason for the former can be explained by the additional freedom in P to model nonharmonic phase behaviour
in comparison to H.
The reason for the latter is that in some cases P detected
double, triple or quadruple period compared to H. However, some of these
periods were too long and were rejected, see for example the case presented in A.3. 
Our second general observation is that, when both P and H have detected a cycle, 
they mostly agree well.
One exception to this is HD103095, where P detected three times longer cycle compared to H. The reasons
here are the same as just explained.
In this particular case the result is obviously physically meaningless as one would not expect the 
phase behaviour of the
signal to contain many extrema. Therefore the true cycle period is most likely around the shorter one obtained by H.

When comparing the results of QP to the other methods we see more dramatic differences. There are several
cases where QP detected significantly different cycle period than H or P. 
The example of HD114710, presented in A.2., illustrates one such case.
Taking into account the fact that quasi-periodic phenomena occur very naturally
in highly non-linear systems, and also 
relying on the results with synthetic
data we would expect the QP estimate to be closer to the true cycle length compared to H and P.

There are several stars for which QP did not detect a cycle while one of the other methods did. 
Mostly this is the case when only P detected the cycle and is therefore explained by the
nonharmonic phase behaviour of the signal, which the regularised QP model was not able to detect.

We note that 15 of the cycles detected using the QP model became significant w.r.t. 
the red noise model. This is not directly an indication that these cycles are more 
reliable, as obviously even some of the visually strongest cycles do not fall into that 
category. However, it might be an indication that on top of the cyclic behaviour there 
is less irregular behaviour for these stars.

In the penultimate column of Tables~\ref{cycles_1} and \ref{cycles_2} we give the
measure of nonharmonicity $\sigma_P$ of the cycle for the QP models.
This quantity is totally unrelated to the error estimate of the cycle period.
Closer look reveals that the values of $\sigma_P$ are rather small due to relatively large time-scale
estimates of the models. Interestingly there are couple of stars for which $\sigma_P$ is 
practically zero, meaning that the cycle is almost harmonic, however, the H model has not
detected the cycle. This is obviously due to the difference in the model selection algorithm. 
The largest value of $\sigma_P$ (also when compared to the cycle period) is seen for HD224930. 
While this can potentially be an example of rather nonharmonic cycle, the 
cycle period estimate is also extremely long, thus the conclusion cannot be strong.

As a validation we should pay attention to the results obtained for the Sun.
We see that the cycle estimates in this case agree quite well with the commonly accepted value 
11 yrs, which is established from the longer datasets. 
The estimate from QP seems to be slightly lower, but still within reasonable 
bounds from the expected value given the error estimate. For the comparison of the fits see Fig. A.1.
in the Appendix.

As a last remark we note that for two of the stars (HD29317 and HD60522) the harmonic
model detected a significant period, but in these cases, due to poor sampling, the highest peaks in the 
spectra were extremely wide and flat.
We have listed these values with question marks in the table and omitted them from the clustering analysis. 
For the former of the objects, QP cycle estimate is significantly higher than that from H 
while P detected roughly two times longer cycle, which was rejected due to our selection criteria.
For the latter of the objects, however, the results from GPs match well with the estimate from H.

\begin{figure*}
	\includegraphics[width=1.0\textwidth, trim={1cm, 1.5cm 1cm 2cm}, clip]{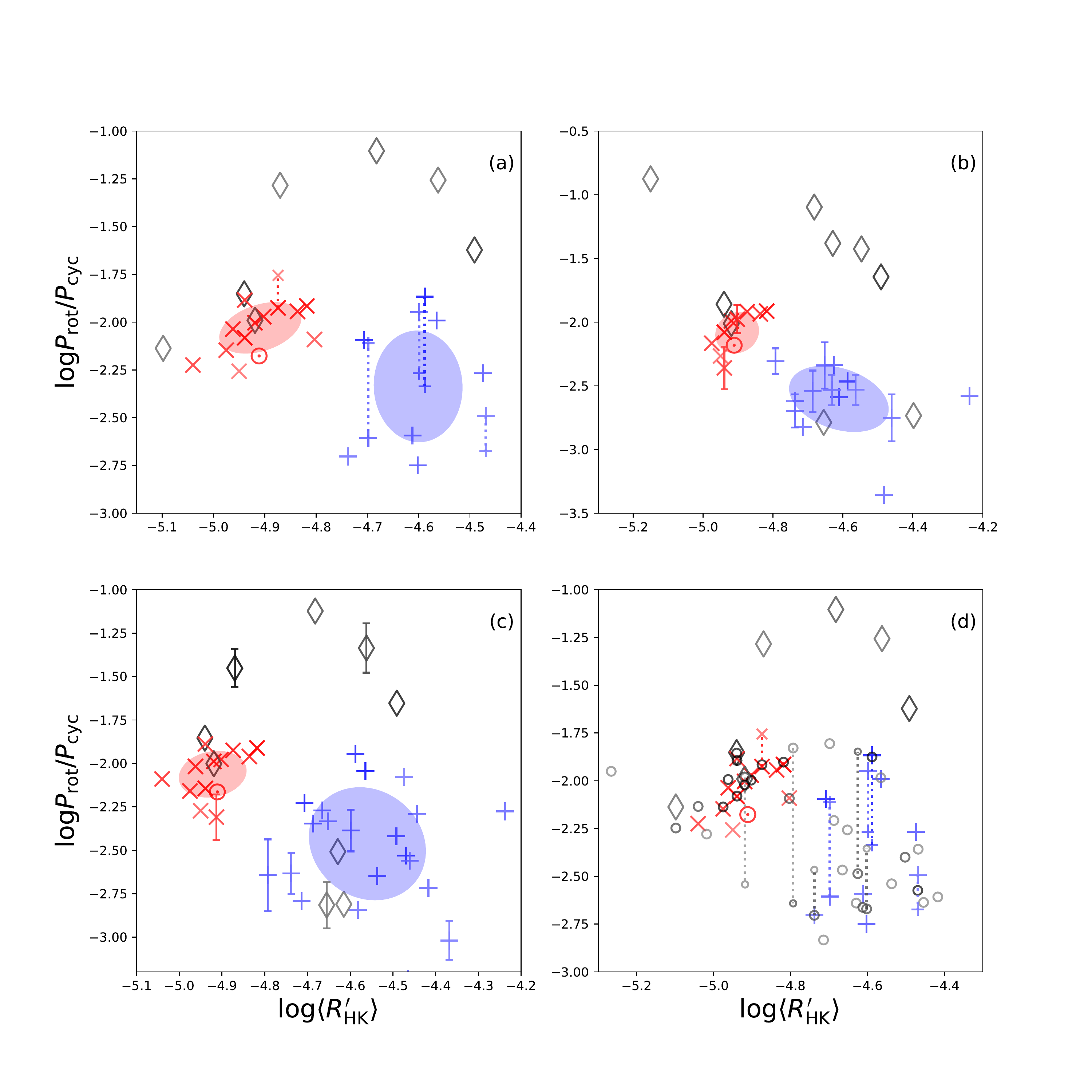}
	\caption{RCRA diagram for cycles obtained from H (a), P (b), QP (c) models and the comparison of the results of H to Cyc95 (d).
	The colour intensities of the symbols indicate the significance of the cycles and the error bars show the 
	$2\sigma$ uncertainties. For better visualization too short error bars have been omitted.
	The vertical dotted lines connect the cycles found for the same star, 
	while the bigger symbol size denotes the primary cycle.
	The Sun is shown with the conventional symbol. The red crosses (blue pluses) 
	represents the stars belonging to the inactive (active) clusters respectively. 
	The black diamonds correspond to the giant stars.
	The ellipses represent the $2\sigma$ regions of the Gaussians obtained from GMM model.
	The small circles correspond to the results from Cyc95.
	}\label{fig_activity_diagram}
\end{figure*}

\begin{figure}
	\includegraphics[width=0.5\textwidth, trim={0 3cm 0 4cm}, clip]{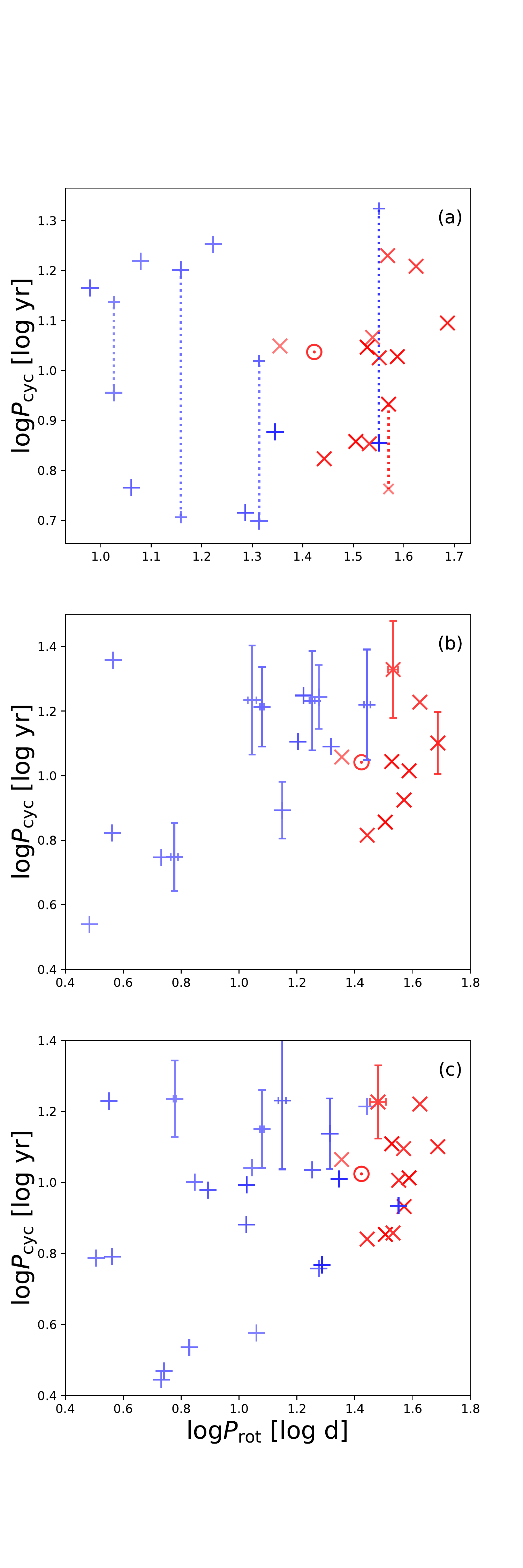}
	\caption{CR diagram for cycles obtained from H (a), P (b) and QP (c) models. The colour 
	and symbol coding is identical to the one in Fig.~\ref{fig_activity_diagram}}
	\label{fig_activity_diagram_2}
\end{figure}

\subsection{Activity diagram}\label{results_activity_diagram}
Next we use the obtained cycle estimates $P_{\rm cyc}$, rotational periods $P_{\rm rot}$ and average 
chromospheric activity indices \RHK\,
to plot the RCRA diagrams \citep[][for a more detailed interpretation see Sect. 5]{BST98,SB99}.
In Fig.~\ref{fig_activity_diagram} we plot three different diagrams separately
for the different methods used to estimate the cycle lengths; panels (a), (b), and
(c) correspond to the results from H, P and QP respectively.
In panel (d) we present a comparison of results from H with the results from Cyc95.
The points in these diagrams, rather independent of the method used,
form two distinct clouds, where the inactive stars locate at somewhat
higher rotation to cycle period values than the active ones; hence a `gap'
in between the clouds, not related to VPG discussed 
before.

Using the points in the activity diagrams as input we further performed clustering 
analysis using GMM with expectation 
maximization algorithm \citep[Chapter~20.3]{BarberBRML2012}.
For simplicity we assumed the points to have no measurement errors and all being equally likely (i.e. we neglected the uncertainty and significance information).
We optimized over the number of clusters, cluster centres and covariances, and selected the model with the 
lowest value of BIC.
We tried models with number of clusters from two to five
and in all three cases the best model turned out to be the one with two components. 
The ellipses in
Fig.~\ref{fig_activity_diagram}
correspond to $2\sigma$ regions of the Gaussians, blue ones to the active cluster and red ones to the inactive cluster. 
With this model, every star is assigned a probability of belonging to either one of the clusters. We have coded the points accordingly to blue 
pluses (red crosses) if the probability of belonging to the active cluster is greater (smaller) than belonging to the inactive cluster. 
Giant stars have not been taken into account in the clustering.

It is evident that, in the case of all three methods, 
the cluster centres agree rather well with each other. The Gaussian distributions
obtained, however, are rather different for all the methods, H and QP
models being more similar to each other, while P shows somewhat distinct behaviour.
The main difference between the results from P and other models is the much wider scatter of points on vertical axis.
We also see that both P and QP have some cycles detected for stars with $\log$\RHK$> -4.4$, while H has not.

As noted before, in all three cases the locations of the cluster centres coincide quite well, however,
their covariances differ much more significantly.
The values of the slopes with $2\sigma$ errors per cluster and each cycle length
estimation method have been collected to Tab.~\ref{tab_slopes}.
For the inactive branch we see that a positive correlation between
$\log$\RHK\, and $\log P_{\rm rot} / P_{\rm cyc}$ is apparent
with all methods, but the uncertainties are relatively large.
The shapes of the active branch ellipses are much wider, indicative of larger scatter,
while all three methods yield negative correlations. However, the correlations in the
case of H and P are very weak with large uncertainties.
Therefore, we cannot confirm the existence of clear positive linear correlations for
both the clusters, in contrast to previous investigations \citep[e.g.][]{BST98,SB99}.
The inactive branch slope, however, is consistent with the earlier studies.

Moreover, \cite{BS18} recently claimed that the negative slope on the active
branch is an unphysical selection effect arising from the lower and upper limits of the
cycle search interval together with plotting quantities that depend on rotation on
each axis. In their plots, they used the inverse Rossby number in the x-axis, computed
from the \RHK~using the rather poorly known convective turnover time empirically
determined by \cite{Noyes1984}. Our analysis in Fig.~\ref{cyc_vs_act} clearly indicates that 
a selection effect due to the upper bound in the cycle search interval for the active
population, showing an increased probability for the occurrence of trends, is present.
In reality, therefore, the cycles in the active population could be even longer, therefore
causing even smaller values of cycle to rotation period relation the higher \RHK~
the star would have. In this case, the real slope would be even more negative,
and the active population even more distinct from the inactive one.

\begin{table}\caption{The slopes of the regression lines in activity branches}
	\begin{tabular}{c|c|c}
		Method & Inactive cluster & Active cluster \\ \hline
		H & 0.93 $\pm$ 0.58 & -0.06 $\pm$ 0.22 \\
		P & 2.06 $\pm$ 1.41 & -0.14 $\pm$ 0.14 \\
		QP & 1.15 $\pm$ 1.39 & -0.46 $\pm$ 0.15 \\
	\end{tabular}
	\label{tab_slopes}
\end{table}

In Fig.~\ref{fig_activity_diagram}(d) we have plotted the results from 
our H model in comparison to Cyc95.
The colour and symbol coding is identical to that in the other panels of that figure, except
that the small black circles correspond to the results from Cyc95 and
their colour intensities
reflect the grade of the cycle estimates. 
The points with poor cycle 
estimates have been excluded from the plot.
Evidently, the inactive branch estimates agree very well with
each other, and the GP methods employed in this study do not
significantly destroy this agreement. Therefore, the collective behaviour of
the inactive stars appears to be robustly captured with the MW sample
independent of the method used.
For the active stars, however, there are much larger differences in between
the methods, and it is very difficult to characterise any collective behaviour
in the RCRA diagram, except that their magnetic cycle over rotation period
ratio appears robustly distinct from the inactive population.
We also see that some of the stars with double cycles from Cyc95 split up between active and inactive
populations \citep[see also][]{Brandenburg2017}, while this does not happen at all based on our results.
This naturally relates to our other conclusions of double cycles themselves
being rare (see our discussion in Sect.~\ref{results_comp_cyc95}).

In \citet{Brandenburg2017} the dependence between $P_{\rm rot} / P_{\rm cyc}$ and
metallicity [Fe/H] as well as relative convection zone depth $d/R$ was investigated by
first calculating the residuals $\Delta_i$ of the data points in the activity diagram
to the linear trend in each branch $i$ ($i$ indicating the active or inactive branch),
to remove the dependency on \RHK\, before correlating with the other quantities.
Although the significance of the linear 
trends based on the results in the current study is rather low, we repeated this procedure,
as the inactive star trend persists in our results.
We used our H model results, but did not find any apparent dependencies with respect to
metallicity nor to the convection zone depth, except that the inactive population showed considerably
smaller residues with less scatter around the mean than the active population.

In Fig.~\ref{fig_activity_diagram_2} we
plot the corresponding CR diagrams for the different methods, with panels (a), (b), and (c) corresponding
to the results from H, P and QP, respectively. The colours of the data
points reflect the cluster labels according to Fig.~\ref{fig_activity_diagram}.
Based on the Cyc95 data, previously \citep[see e.g.][]{BohmVitense2007} linear trends were
detected in such diagrams for both populations, and the Sun was quite clearly located in between
these trends as a solitude object \citep[see e.g.][]{BohmVitense2007}. 
The inactive population stars exhibit positive linear $P_{\rm rot}$ -- $P_{\rm cyc}$ correlation, that is, the faster the rotation, the 
shorter the cycle, with all methods used, but the scatter around this
trend is far more pronounced as seen in Cyc95 data by \cite{BohmVitense2007}.
The Sun, however, is no longer far off from the common trend, and is no longer a solitude
object. 
The cycle lengths of the active stars show no trend in this diagram with H and QP methods, while
a hint of a linear trend similar to that claimed by \cite{BohmVitense2007} can be seen with method P.

\subsection{Giant stars}
The MW sample contains 54 giant stars for which more than ten years of data is available, and which 
are thus included in our analysis. For 17 of them we detected cycles with at least
one of our methods. Therefore, the presence of cycles within the giants is less
likely (31 \% show cycles) than in the MS stars (56 \% show cycles).
This is not surprising, as the rotation periods of these stars are generally much longer, and
therefore one would assume that the magnetic activity level would be reduced due to a less
efficient dynamo, or perhaps a large-scale dynamo would not be excited at all, in which case
no cycles would be detected. 
Remarkably, we detect fairly significant magnetic cycles for four giant stars that have rotation
periods exceeding 100 days (HD10072, HD29317 and HD68290) and even 200 days (HD124897),
all with chromospheric activity indices matching with the MS stars.

Based on Fig.~\ref{fig_activity_diagram} we can see that roughly half of the cyclic giant stars
fall on the upper part of the RCRA diagram, that is, clearly above both of the inactive and active
population clusters, indicative of relatively short magnetic cycles in them.
This could, however, be an observational bias due to the even more limited extent of the
data set with respect to the rotation periods. For example,
the star HD68290 has a rotation period of 158 days, which means
that only a cycle length longer than 40 years would place the star
on any of the activity clusters on the diagram. 
Indeed, some of the giant stars with shorter rotation periods,
although somewhat depending on the analysis method,
fall into the inactive and active populations. 
It is nevertheless interesting
to note that the MS stars seem to have an upper limit of the rotation to cycle period
ratio values at around -1.8, although 
higher values
could technically have been observed from our sample, while at the same time
there are giant stars that have cycles similar to those of the MS stars.

\section{Discussion in the light of theory and numerical models}\label{theory}

\subsection{On the Vaughan-Preston gap}

There are many dynamo-related explanations postulated to lead to the VPG.
Some involve arguments of the dynamo mode changing its topological
complexity as function of rotation, due to the changing supercriticality
of the dynamo solution, from simple to complex ones
causing a drastically different chromospheric response \citep[e.g.][]{Durney1981}.
Others postulate that even the location of the dynamo might change
from a near-surface one for the active stars to one operating near the
bottom of the convection zone for the inactive ones \citep{BohmVitense2007}.
\citet{Metcalfe2016} proposed that the operation of dynamos could be disrupted due to
dramatic changes in the differential rotation, the rotation law changing
from solar-like to antisolar one. Also enhanced magnetic braking, leading
to a decreased probability to detect stars around certain \RHK\, values,
has been proposed \citep[e.g.][]{Noyes1984}.
If such drastic changes would actually occur, one would expect to
see abrupt changes in the quantities describing the large-scale
dynamo at around $\log$\RHK$=-4.8$ corresponding to the VPG.

Both the RCRA and CR diagrams reveal changes in the cycle length
behaviour at around $\log$\RHK$=-4.8$, and the re-estimation of the cycle
lengths compared to Cyc95 makes this distinction even clearer: in the activity
diagrams presented in this study hardly no overlap of cycle lengths w.r.t. \RHK\, or
rotation period occurs in between the active and inactive populations, while
some overlap was reported in earlier studies \citep[e.g.][]{BST98,SB99,BohmVitense2007},
mainly in terms of primary and secondary cycles, if detected, falling on different branches.
Intriguingly, however, the histogram of the C
class of stars
does not show the prominent bimodal distribution
of the whole MW sample as seen from Fig.~\ref{cyc_vs_act}(b).
This indicates that
the tendency for cyclic behaviour, a sign of a large-scale dynamo in
action, is not similarly reduced as the
total distribution of stars in the VPG.
The distribution of the relative variance of C stars is similar, that is,
unaffected by the VPG.
Hence, the data is inconsistent with a disruption of large-scale
dynamo action in the VPG, but possibly consistent with a smooth dynamo mode
change affecting the cycle lengths but not the overall
efficiency of the large-scale dynamo.
Moreover, the existence and even slight overabundance of C stars in the VPG is clearly against any
scenario that relies on enhanced temporal evolution due to
efficient braking by ordered magnetic fields.

Interestingly, within the inactive population, we observe an abrupt change in the C star 
variances  at about the solar activity index $\log$\RHK$\approx -4.9$. 
This could
be an indication of 
an abrupt change in the operation of the large-scale dynamo.
A prominent candidate for such a transition is the change
from solar to anti-solar rotation profiles.
Both numerical models \citep[e.g.][]{Brun2017,Viviani2017}
and also a recent study where observations were interpreted with the
help of numerical models \citep{Giampapa2018} support such a
transition near the solar parameter values.

\subsection{Cycle lengths in different populations}

\cite{BohmVitense2007}, in an attempt to explain the linear
trends seen in the CR diagrams of Cyc95 cycles,
proposed that the dynamo type would abruptly change in between the two
populations, from an interface dynamo working
near the surface (for active stars) to one working at the bottom of their
convection zones with deeper mixing (for inactive stars). Hence, one would also
expect a dependence of the cycle lengths and properties on the convection zone 
depth.
\citet{Brandenburg2017} tried
to find such a dependence, essentially using the Cyc95 cycles, without
success. Neither do we after our re-analysis. Therefore, the different
cycle lengths are unlikely to reflect such a drastic change in the operation
of the dynamo.

\begin{figure}
	\includegraphics[width=0.5\textwidth, trim={0 0cm 1cm 1cm}, clip]{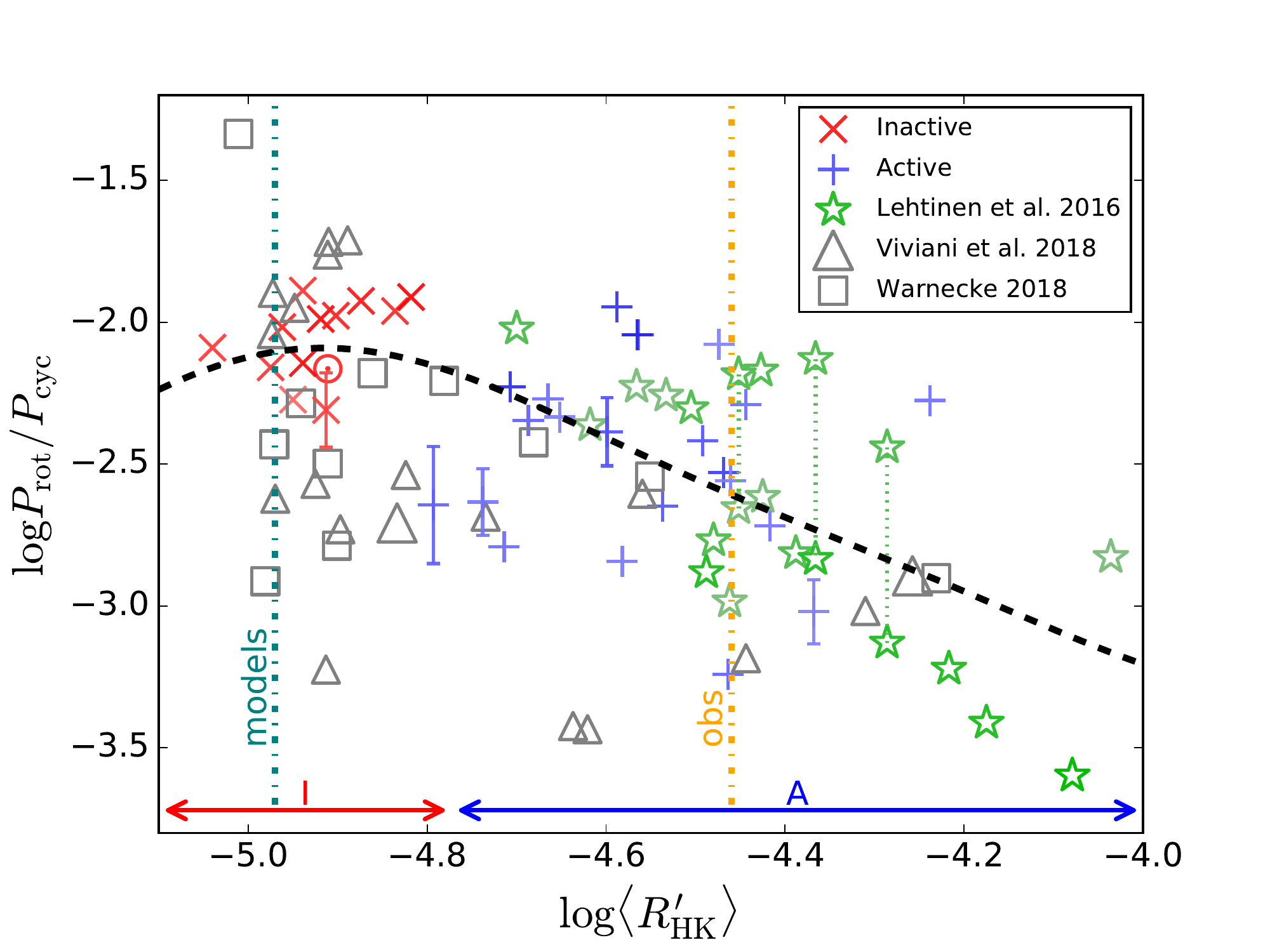}
	\caption{The comparison of the results of quasi-periodic GP model and different other
                observational and modelling studies. 
		Green stars
                are taken from the analysis of long-term photometry of solar-type stars
                \citep{Lehtinen2016}, where intensity reflects the cycle grade.
                Grey squares are from the wedge study of turbulent magnetoconvetion
                by \citet{Warnecke2017b}, while
		grey triangles are from a corresponding study of \citep{Viviani2017},
                where the wedge assumption was relaxed in the azimuthal direction, allowing for
                nonaxisymmetric solutions. Larger triangles indicate high-resolution runs.
                In the bottom of the figure we have shown the ranges of the inactive, active 
                and transitional branches using arrows.
                The vertical dash-dotted lines mark the points where axi- to nonaxisymmetrical transition
                occurs accordingly to models and observations.
                The dashed black curve represents a blend of two linear fits to the inactive 
                and active plus transitional branch.
                The remaining colour and symbol codings are identical to the one in Fig.~\ref{fig_activity_diagram}.
                Here we have used the magnetic to kinetic energy ratio obtained in the models as a proxy
                for calculating the \RHK\, activity measure. The runs with solar rotation rates have been
                fixed to the solar \RHK\, value, and to scale the axis we have used the relation 
                \RHK $\propto \sqrt{\overline{B}}$.
                Two of the simulations from \cite{Viviani2017} with highest magnetic activity levels, being incomparable
                with observational data, have been omitted from the plot.
	}\label{fig_activity_diagram_cmp2}
\end{figure}

In the light of more recent results, it is likely that a different
type of dynamo transition occurs in between these two populations.
The change from nonaxisymmetric to axisymmetric dynamo modes is
a potential candidate to occur close to the
transition between the branches, but the current observational
data from photometry suggests that the transition points do not exactly match,
the nonaxi- to axisymmetric point being located at higher chromospheric activity indices,
compare the location of the two vertical lines in Fig.~\ref{fig_activity_diagram_cmp2}.
Numerical models
place this transition to lower values of the activity index, but
relating the models with the observables is
still quite difficult, as the models are very likely too laminar.

On the other hand, the absolute differential rotation, significant for the dynamo action,
remains constant or is at best weakly varying
as function of rotation \citep[see the summary of observational results in][]{Lehtinen2016}.
The role of turbulent effects, in contrast, is theoretically expected 
to grow as function of rotation, at least
in the regime of slow rotation. This implies
a change from $\alpha \Omega$ dynamos into the regime where both
contributions are equally important ($\alpha^2 \Omega$). Those global
magnetoconvection models that
show transition to nonaxisymmetric magnetic field configurations also
show that the differential rotation becomes strongly quenched at rapid
rotation \citep{Viviani2017}, which might even imply that the dynamos
in the very active stars could be of $\alpha^2$ type.

From this it follows that it is not completely safe to assume that the
results obtained for the well-studied axisymmetric solar-type $\alpha \Omega$
dynamos would directly apply to the stellar populations studied here.
A similar wealth of studies of nonaxisymmetric $\alpha^2 \Omega$ or $\alpha^2$ dynamos, 
unfortunately, does not exist. This relates to the difficulty to obtain oscillatory
$\alpha^2$ solutions, even though they have been shown to exist
\citep[see][and the references therein]{Brandenburg2017a2}. Some useful, albeit
simple, studies have been undertaken, however. 
\citet{Cole2016} studied the cycle frequency in $\alpha^2\Omega$ dynamos with varying amounts of
differential rotation in one-dimensional kinematic dynamo models
in spherical geometry. They
found out that the general behaviour of such dynamos is similar to their
$\alpha \Omega$ counterparts in that the cycle period is a decreasing
function of increasing amount of differential rotation. With strong shear,
the cycle period relation directly following from the dispersion relation
for $\alpha \Omega$ dynamos, Eq.~(\ref{alpom}), was followed, but when the strength
of differential rotation was decreased, they observed rather a nonmonotonous
behaviour with a regime, where the cycle period was strongly increasing when
$\Delta \Omega$ was decreased only a little. When $\Delta \Omega$ was
decreased even further, another power law was established
with considerably longer cycle lengths. The nonmonotonous behaviour was linked with
a change in the dynamo solution topology, that is, antisymmetric (with respect to
the equator) solar-like solutions changed into symmetric ones.
A similar, but less pronounced jump in cycle periods from short to long ones was observed close
to the transition from axi- to nonaxisymmetric dynamos
by \citet{Viviani2017}.
Studying these types of dynamos further is, therefore, one potential track to
understand the cycle length distributions in the two populations.

\subsection{Rotation period to cycle length ratio}

In the data analysis section we concluded that 
the physically most plausible model is the quasi-periodic one, best suitable
for modelling highly supercritical nonlinear oscillators, such as the Sun.
Hence, we plot the results from our QP models together with some previous ones from observations \citep{Lehtinen2016},
and also from models \citep{Warnecke2017b, Viviani2017} in an RCRA diagram in Fig.~\ref{fig_activity_diagram_cmp2}.
Especially the observational points support the interpretation of there being only
two branches (inactive and transitional) instead of three (inactive, active and transitional).
The inactive branch has a positive slope, while the active branch appears to
merge with the transitional branch, which together exhibit a steep negative
slope. 
The main scatter in this plot is due to the points taken from global
magnetoconvection models.

Also in the CR diagrams we see only one clear trend, that is the decreasing magnetic
cycle length as function of rotation rate for the inactive population.
Already this single trend constrains
the dynamo solutions, as the two prevailing dynamo paradigms give clearly
opposite predictions for it: flux-transport dynamos, accepted as the
standard model for a solar dynamo, predict a dependence such that the
cycle length grows with rotation, that would manifest as a negative
slope in CR-diagrams (opposite to what is observed) \citep[e.g.][]{Jouve2010}.
The standard $\alpha \Omega$ dynamo model based on helical turbulence and
differential rotation acting together throughout the convection zone would give
a cycle period
\begin{equation}
P_{\rm cyc} \propto \left| \alpha \Delta \Omega \right|^{-1/2}, \label{alpom}
\end{equation}
where $\alpha$ and $\Delta \Omega$ describe the strengths of the
inductive effects arising from turbulence and radial differential rotation,
respectively \citep{Stix1976}.
This formula predicts $P_{\rm rot} / P_{\rm cyc}$ to grow with $P_{\rm rot}$
as the turbulent effects are expected to grow in the regime of slow
rotation proportional to $\Omega$ \citep[e.g.][]{KR80}.
So, the trend predicted by the turbulent dynamo is of correct sign, but it
appears to be far too steep to be easily explained. 

This can be seen by comparing
the expected scalings in a RCRA diagram from a kinematic $\alpha \Omega$ dynamo,
that predicts the rotation period to cycle length ratio to scale as
\begin{equation}
P_{\rm rot}/P_{\rm cyc} \propto \Omega^{-1} \left| \alpha \Delta \Omega \right|^{1/2}. \label{pcyc}
\end{equation}
If we make the following ansatz for the dependencies
for the relevant effects, $\alpha=\Omega^a;\ \Delta \Omega = \Omega^b$, we
obtain the following relation from Eq.~(\ref{pcyc}):
\begin{equation}
a+b=2\left(\nu+1\right),
\end{equation}
with the slope deductible from observations, $\nu$.
This immediately demonstrates the evident: values of $\nu$ would need to become negative to bring
the scaling of $\alpha$ to acceptable limits ($a$ maximally 1), with values of $b$ agreeing
with observations and theoretical arguments $(0 \dots 1)$.
This study, using the acceptable limits of $b$ stated above with the $\nu$
values from the QP method, gives 
values of $(a \approx 3.3 \dots 4.3)$, far too high to be realistic, agreeing with
previous studies.

These trends were explained by \citet{BST98} by assuming that the dynamo coefficients are
increasing functions of the magnetic field, based on the known dependence of \RHK\, on the magnetic field. 
This appears as unlikely behaviour for the convection-driven $\alpha$-effect
\citep[see e.g.][]{Karak2014}.
More likely scenarios include an $\alpha$-effect being
driven by some other instabilities, such as MRI \citep[e.g.][]{Masada2011},
buoyancy instability \citep[e.g.][]{Chatterjee2011}, or
the shear-current effect feeding from the small-scale magnetic field
\citep[see e.g.][and references therein]{Squire2016}. Such effects, however, are
only beginning to be considered in stellar dynamos, even though the obvious need,
as neither of the prevailing dynamo paradigms offer an explanation to the
inactive branch scaling.

Nowadays an abundance of global magnetoconvection models are available,
that allow for the direct modelling of the stellar dynamo mechanism, and
recently studies, where different rotation rates have been considered, have emerged
\citep{Strugarek2017,Warnecke2017b,Viviani2017}. Those models that produce 
mostly axisymmetric dynamo solutions either by the design of the model itself
\citep[e.g.][see also rectangular points plotted in 
Fig.~\ref{fig_activity_diagram_cmp2}]{Warnecke2017b} 
or self-consistently \citep{Strugarek2017} tend to agree best with the negative
slope for the active branch or with the even steeper slope obtained
for even more active stars \citep[the transitional branch, e.g.][]{Lehtinen2016}, while
no clear inactive population emerges. Remarkably, many of the properties
of the obtained dynamo solutions can be explained with the simple $\alpha \Omega$ 
cycle period scaling,
Eq.~(\ref{pcyc}), when the turbulent transport coefficients are directly measured 
from the models \citep[see e.g.][]{Warnecke2014,Warnecke2017,Warnecke2017b}.
The models of \cite{Viviani2017} allow for and also excite nonaxisymmetric solutions
(see the triangular points in Fig.~\ref{fig_activity_diagram_cmp2}). In this case, 
two distinct populations emerge, but neither of them are in very good agreement with
observations: the inactive branch does not well co-incide with the observed one, although
shows hints of positive slope, while the transitional and superactive branches seem
to have merged into one single population with a much shallower negative slope than
observed. Although this seems as a very promising approach, it has to be borne in mind
that these global magnetoconvection models are unrealistic also in the sense that they 
produce dynamos that behave very much solar-like, but the cause for this behaviour 
is not the same as in the Sun \citep[][]{Warnecke2014}. In contrast, 
the required change of sign of the $\alpha$ with respect to $\Omega$ effect to obtain
equatorward migration results from a region of negative shear present in the models, but
such regions are not observed in the Sun. 

Last we note that there would also be a dependence
on the square root of the length scale in Eq.~(\ref{alpom}), 
which was neglected here and previously, as
it appears natural to assume a typical length scale of the dynamo would be of the order
of the depth of the convection zone, and this
parameter would not change as function of rotation within a given spectral type.
The very recent results obtained from turbulent convection modelling, however,
have revealed the emergence of sub-adiabatic (formally convectively stable) layers
within what is normally considered the convection zone
\citep[see e.g.][and references therein]{Kapyla2017}, possibly changing the location
and extent of the dynamo-active layer, which could also depend critically
on the rotation rate.
Even a weak dependence of the length scales on rotation
(such that the relevant length scale grows with rotation) could render the classical dynamo
compatible with the observations. One such scenario arises as follows: given a critical
Rossby (or Coriolis) number at which the dynamo turns on, the depth of the dynamo-active
part of the convection zone increases with decreasing (increasing) Ro (Co) inversely (directly)
proportional to the rotation rate. If such convectively stable layers existed in reality,
they could act as a storage of magnetic flux, playing at least partly a similar role
than tachoclines are supposed to do for solar dynamos. 
Such layers have been recently detected in global convection models
\citep[][]{KMB18,KramersGlobal18}, and the dynamo solutions were indeed found
to be sensitive to the changes in the convection zone structure \citep{KramersGlobal18}.
Also, such layers might critically contribute to reversing the sign of helicity of the
flow, a frequently observed phenomenon in convection models with an overshoot
layer \citep[see e.g.][]{Kapyla2004}, that could help in getting the remaining
details correct in the turbulent magnetoconvection models \citep{Duarte2016}.
Such a helicity inversion has already been reported to occur in the global
magnetoconvection models \citep{KramersGlobal18}, but its effect appeared rather
subtle in these models.

\section{Conclusions}\label{conc}
In this paper we have presented a re-analysis of the MW chromospheric activity sample starting with
a harmonic model with a trend, and refining the model towards a more realistic situation
where we allow for quasi-periodic cycles, that highly nonlinear systems often produce.
We have identified several potential sources for erroneous detections of periods in the
case where the data set length is of the same order of magnitude as the cycles that are
searched, and in addition the sampling is uneven containing large gaps.
These include the improper treatment of linear trends and possibly too simple assumptions
of the noise variance model. Assumption of strict harmonicity can result in the appearance
of double cyclicities that seem more likely to be a result of the quasi-periodicity of the
cycles. Consequently, we conclude that only rare cases of reliable double cycle detections
can be made based on the MW sample.

We observe an increased tendency of the active population
to show trends that are unlikely to be instrumental artefacts. This is indicative of
the presence of longer than detectable cycles in them. What is detectable from the
MW sample could be sub-dominant secondary shorter cycles, while the longer basic ones
could remain undetected. In conclusion, the MW sample cannot
be regarded to well represent this population's magnetic cycles.

We observe pronounced cycles also in stars that have \RHK\, activity indices
belonging to the so called VPG. 
The distribution of all
stars, however, shows a clear gap. This suggests that the gap is not related
to the operation of the large-scale stellar dynamo. At around solar \RHK\,
values and slightly below, however, we see a clear drop in year-to-year variance of the cyclic
stars, indicating a disruption of cyclic dynamos there. Such a disruption could be
caused by the transition of solar-like differential rotation profiles into
anti-solar ones, at present indicated by many numerical models. It
remains, however, rather unclear how exactly should these models be scaled back
to the real stars to allow for direct comparisons.

We confirm the earlier claims of the existence of two clearly distinct active and inactive
populations in the MW sample, based on a clustering analysis performed with GMM. 
We also confirm the claim of the inactive stars to show a clear linear trend
with a positive slope in the RCRA diagram, while discard the claims
of the active stars showing a similar trend \citep[see e.g.][]{BST98,SB99}.
The data is consistent with the active population representing the less active
tail of the transitional branch stars, analysed from other data sets
for example by \citet{Lehtinen2016,Distefano2017}. 

One interesting question that remains unanswered in this study due
to the high observational noise and insufficient length of the datasets is how coherent the found
cycles are. One could think that the coherency of the cycle would
be dependent on how developed the dynamo of the given star 
is.
Even though we reported the measures of nonharmonicity $\sigma_P$, 
they are only of theoretical interest because of the high uncertainties in time-scale estimates.
Nevertheless, we showed with the tests on synthetic data, 
that the quasi-periodic model can be used to improve the cycle 
period estimates compared to the values obtained from the fully harmonic model. 

In the current study we downsampled the data to make the computations of the GP models feasible.
Another possibility to efficiently model the time series would be to use state space models. It has been shown that GPs 
with periodic and quasi-periodic covariance functions can be reformulated in state space models which reduces the computational 
complexity to linear in number of time steps \citep{Solin2014}. The link in the opposite direction between different state space 
models and GP covariance functions, including the quasi-periodic one is discussed in \citet{Grigorievskiy2016}. To reduce the 
computational cost, instead of downsampling the data, one can also use sparse GP models. 
In these approaches the subset of the inputs and model parameters are 
simultaneously optimized to obtain a good approximation to the full GP model \citep{Titsias2009}. 
Applying the aforementioned methods 
to astronomical datasets we consider as interesting topics for the future studies.

\begin{acknowledgements}
  This work has been supported by the Academy of Finland Centre of
  Excellence ReSoLVE (NO, MJK, JP), Finnish Cultural Foundation grant no. 00170789 (NO) and 
  Estonian Research Council (Grant IUT40-1; JP).
  We thank Aki Vehtari for useful discussion and comments on GP related questions. MJK and JL
  gratefully acknowledge discussions with Petri K\"apyl\"a, Andreas Lagg, and J\"orn Warnecke
  on the theoretical interpretation of the observational data.
  We also thank Katalin Olah for sharing insight into some questions related to MW dataset,
  as well as Matthias Rheinhardt and Axel Brandenburg for comments on the manuscript.
\\
\\
The HK\_Project\_v1995\_NSO data derive from the Mount Wilson Observatory HK Project, which was supported by both public and private funds through the Carnegie Observatories, the Mount Wilson Institute, and the Harvard-Smithsonian Center for Astrophysics starting in 1966 and continuing for over 36 years.  These data are the result of the dedicated work of O. Wilson, A. Vaughan, G. Preston, D. Duncan, S. Baliunas, and many others.  
\end{acknowledgements}

\bibliographystyle{aa}
\bibliography{paper}

\begin{appendix}

\section{Comparison of harmonic, periodic GP and quasi-periodic GP models}\label{example_fits}
In this section we illustrate the differences between the fits of H, P and QP
models using some of the datasets as examples. 
The regression curves for the Bayesian models are calculated as the means of the 
posterior predictive distribution $p(g(t_*)|t_* ,\mathcal{D})=\int_\Theta p(g(t_*)|t_* ,\theta,\mathcal{D})p(\theta|\mathcal{D})d\theta$, where $g(t_*)$ is the function value at test time moment $t_*$, $\mathcal{D}$ is the observed data, $\theta$ are the parameters and $\Theta$ is the domain of $\theta$. In particular 
for the GP models the posterior predictive is a Gaussian with mean 
$\overline{g}(t_*)=\mu + \vect{k}_*\tran \vect{K}^{-1}\vect{y}$ and variance 
${\rm var}(g(t_*))=k(t_*, t_*)-\vect{k}_*\tran\vect{K}^{-1}\vect{k}_*$, where $\vect{k}_*=[k(t_*,t_1), \dots k(t_*,t_N)]\tran$ and the remaining symbols were defined in Sect.~\ref{GP_models}.

\begin{figure}
	\begin{tabular}{c}
		\includegraphics[width=0.5\textwidth, trim={0 0.6cm 0 1cm}, clip]{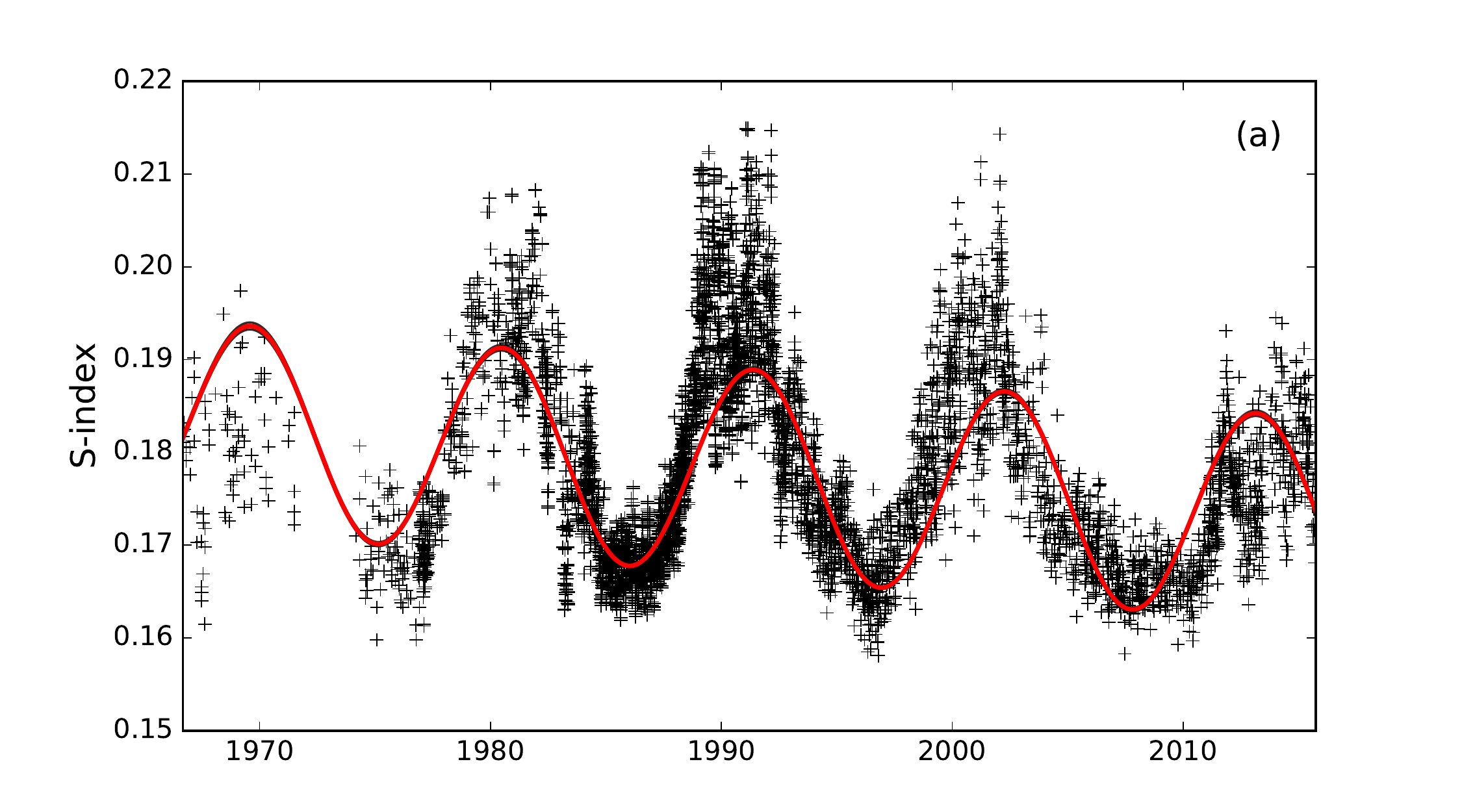} \\
		\includegraphics[width=0.5\textwidth, trim={0 0.6cm 0 1cm}, clip]{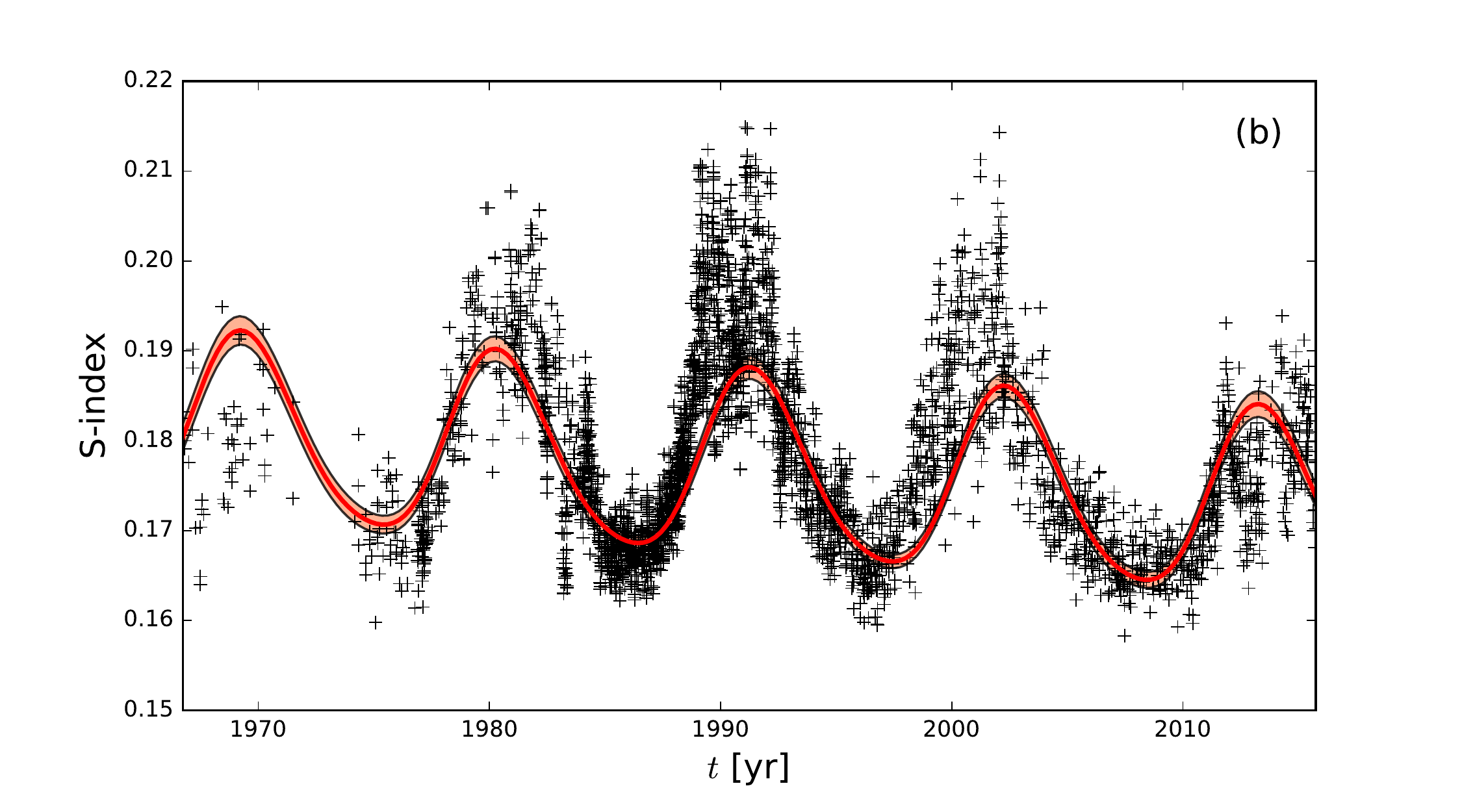} \\
		\includegraphics[width=0.5\textwidth, trim={0 0cm 0 1cm}, clip]{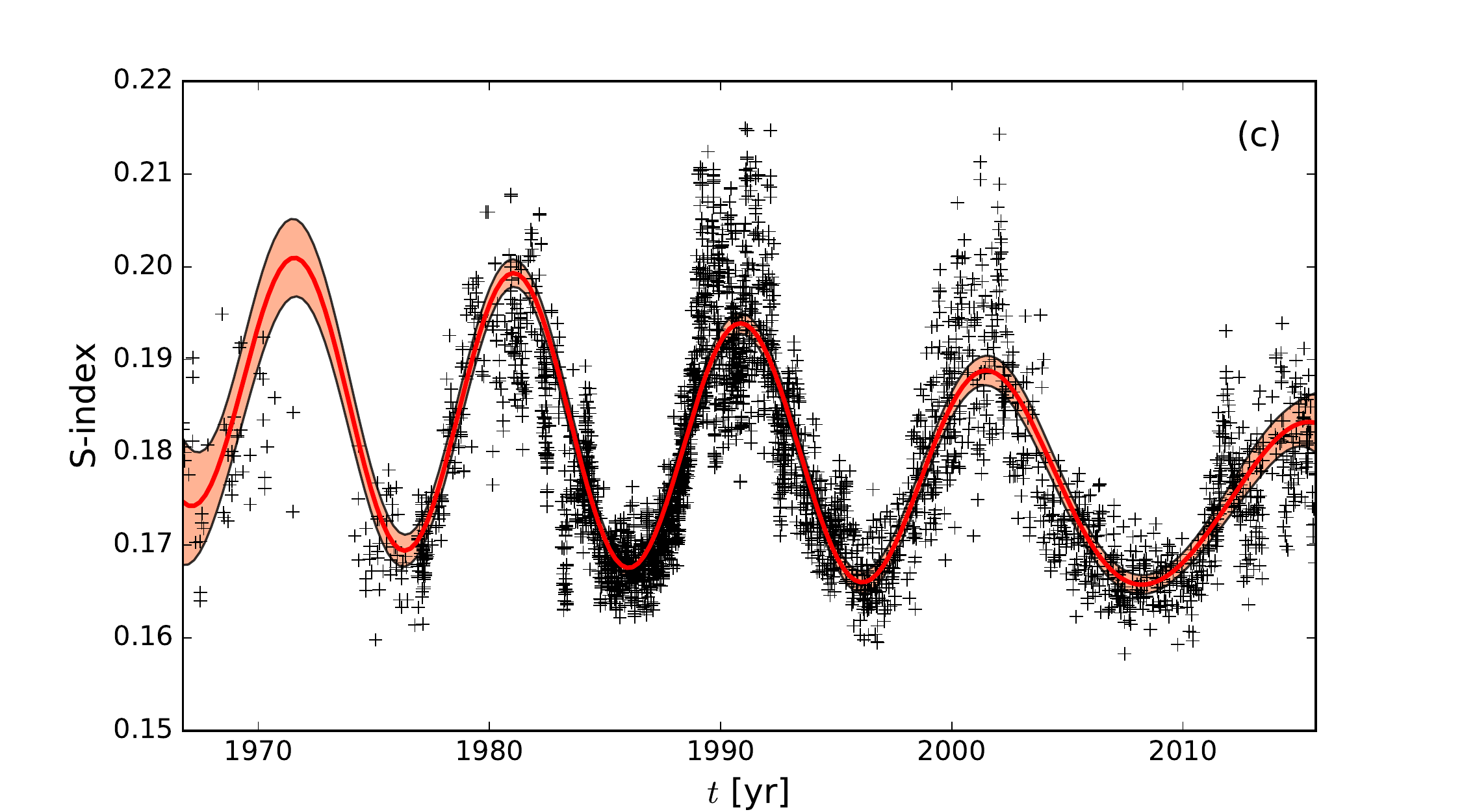}
	\end{tabular}
	\caption{Comparison of the fits of H (a), P (b) and QP (c) 
	models for the Sun. Black crosses are data and red lines the predictive mean curves. The shaded areas 
	around the means of the GPs correspond to 2$\sigma$ intervals of the 
	predictive distributions.}
	\label{fig_sun_fits}
\end{figure}

We start with the results for the Sun which are depicted in Fig.~\ref{fig_sun_fits}. 
On the top panel we have plotted the mean curve of the H model. It is evident that single harmonic is 
not realistic model for the data, as it clearly goes out of phase with the data at different cycles. The 
evidence of the trend in the solution is relatively strong. In the middle panel there is shown a fit from 
P model. In this case the quality of the fit is somewhat better due to more complex form of the 
periodic, however we see the similar tendency of the mean curve being slightly out of phase during all but 
the middle cycle. The linear trend is also evident. On the bottom panel the results from QP model
show much better overall fit to the data. As the time-scale of the model is only 1.5 times longer than the cycle period,
we see quite heavy modulation of the cycle amplitude. The presence of the trend is also apparent in the model.

\begin{figure}
	\begin{tabular}{c}
		\includegraphics[width=0.5\textwidth, trim={0 0.6cm 0 1cm}, clip]{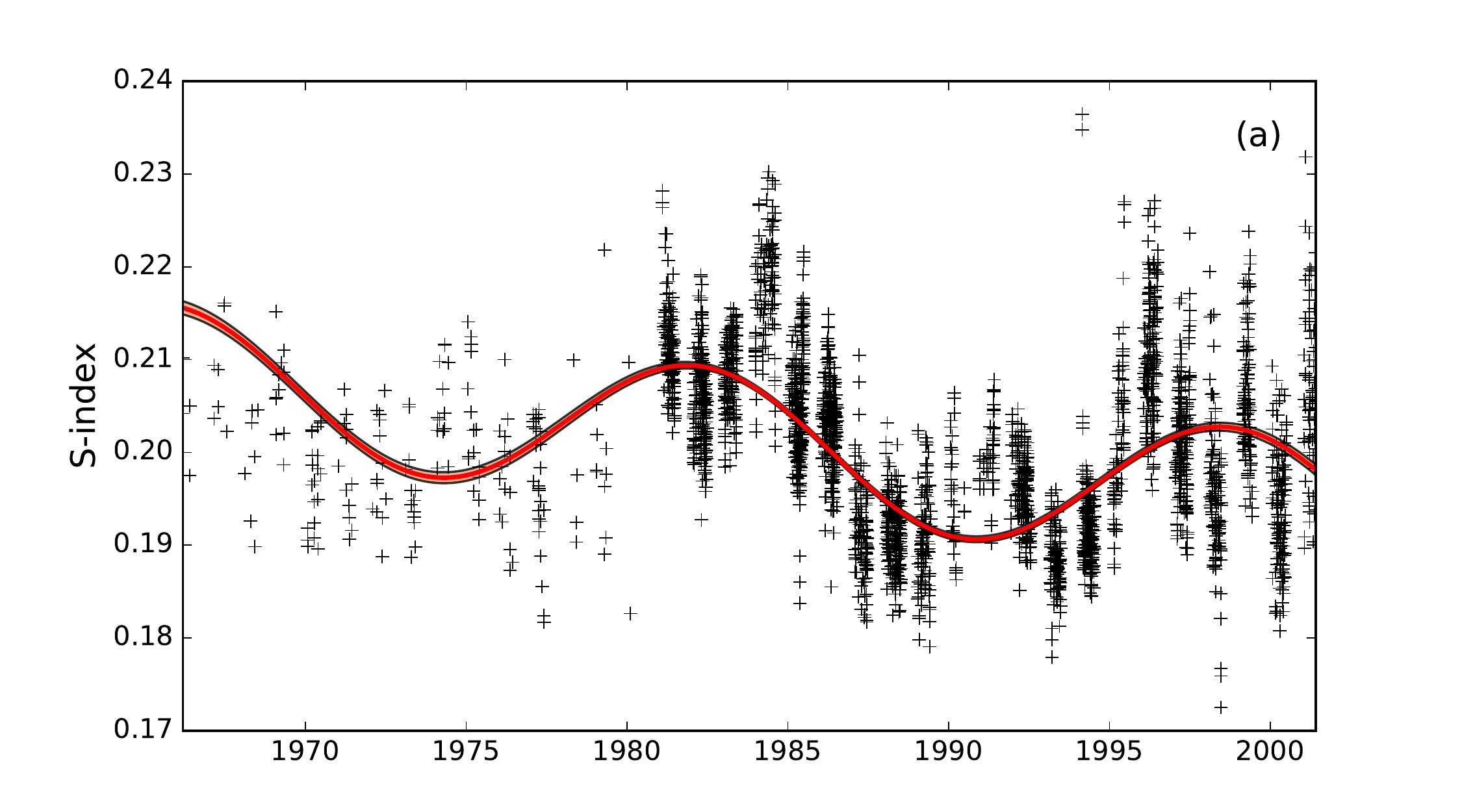} \\
		\includegraphics[width=0.5\textwidth, trim={0 0.6cm 0 1cm}, clip]{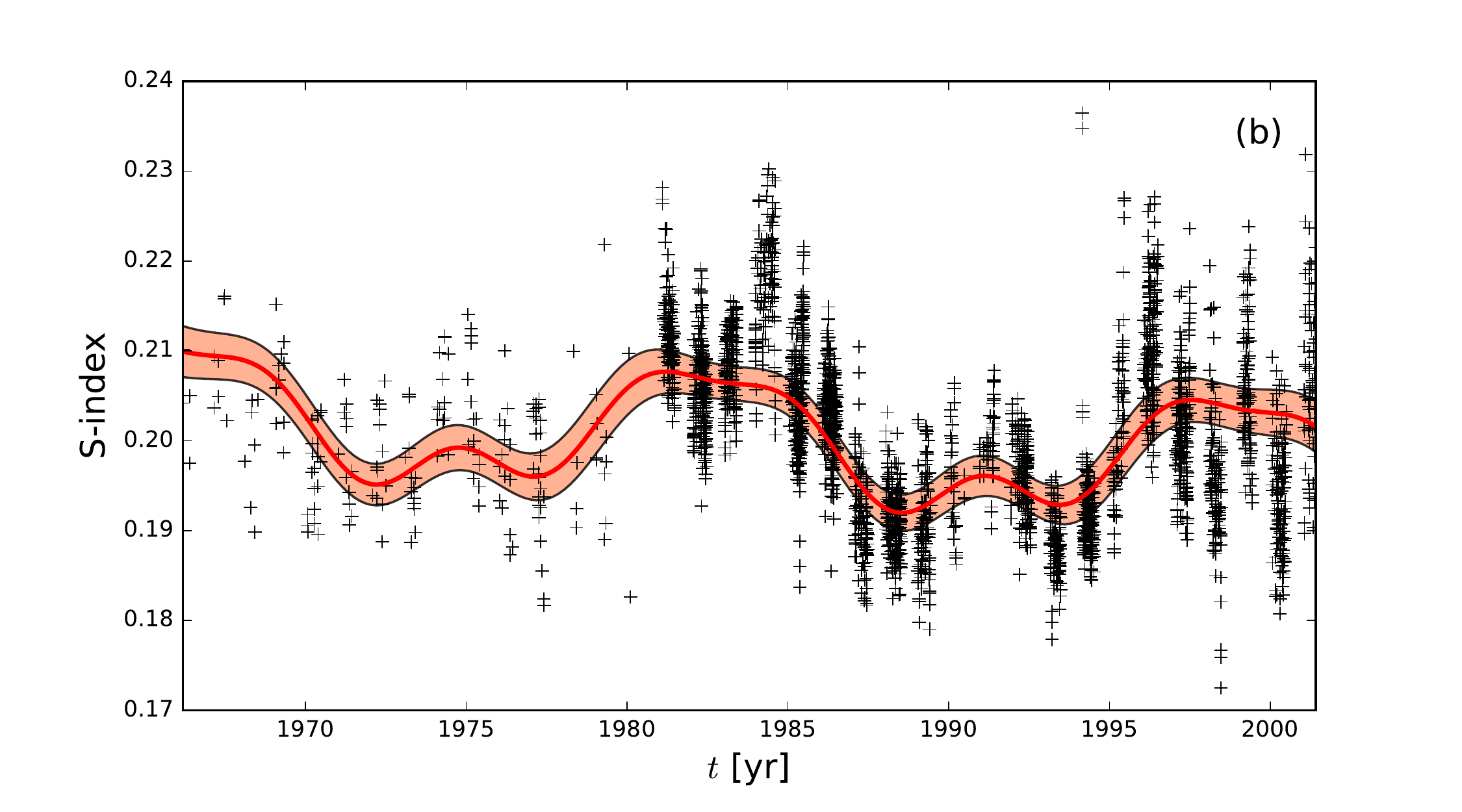} \\
		\includegraphics[width=0.5\textwidth, trim={0 0cm 0 1cm}, clip]{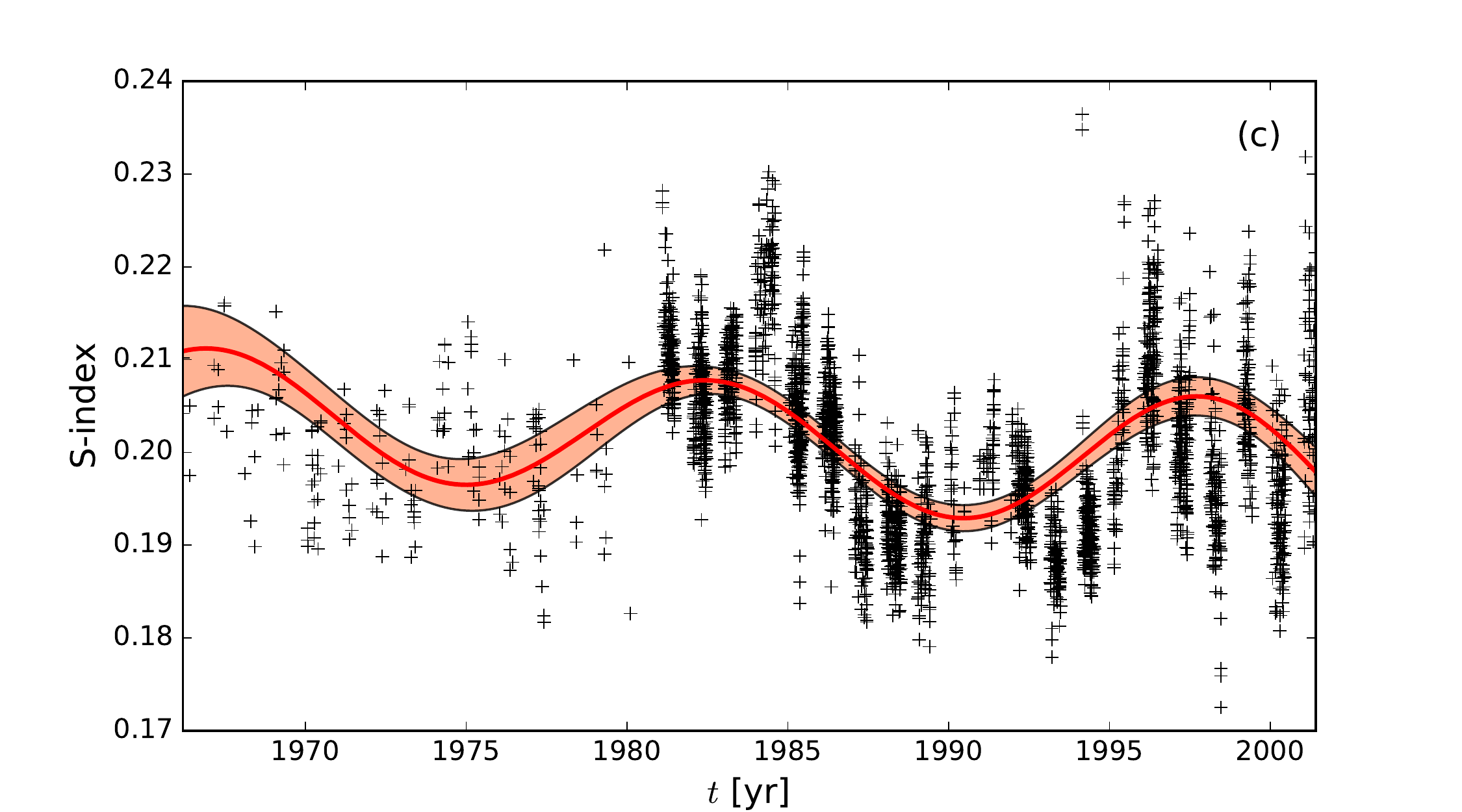}
	\end{tabular}
	\caption{Comparison of the fits of H (a), P (b) and QP (c) 
		models for HD114710. The meaning of the panels and symbols is the same as in Fig.~\ref{fig_sun_fits}.}
	\label{fig_114710_fits}
\end{figure}

The next example is the star HD114710, which we selected because of a significant difference in the period estimate
for the QP compared to the other models. The fitted curves can be seen in Fig.~\ref{fig_114710_fits}. 
From the H model we see that there is a period around 16 yrs. In Cyc95 there was 
detected also a secondary harmonic cycle, which in our analysis, however turned out to be insignificant. 
From the P fit we see that the dataset is still explained by one periodic, which now has the 
phase behaviour containing two maxima. So the potential secondary harmonic cycle is enclosed in the behaviour of a single periodic.
QP fit on the other hand detected the shorter cycle period around 14 yrs. The time-scale of this model is
approximately seven times longer than the cycle period.

\begin{figure}
	\begin{tabular}{c}
		\includegraphics[width=0.5\textwidth, trim={0 0.6cm 0 1cm}, clip]{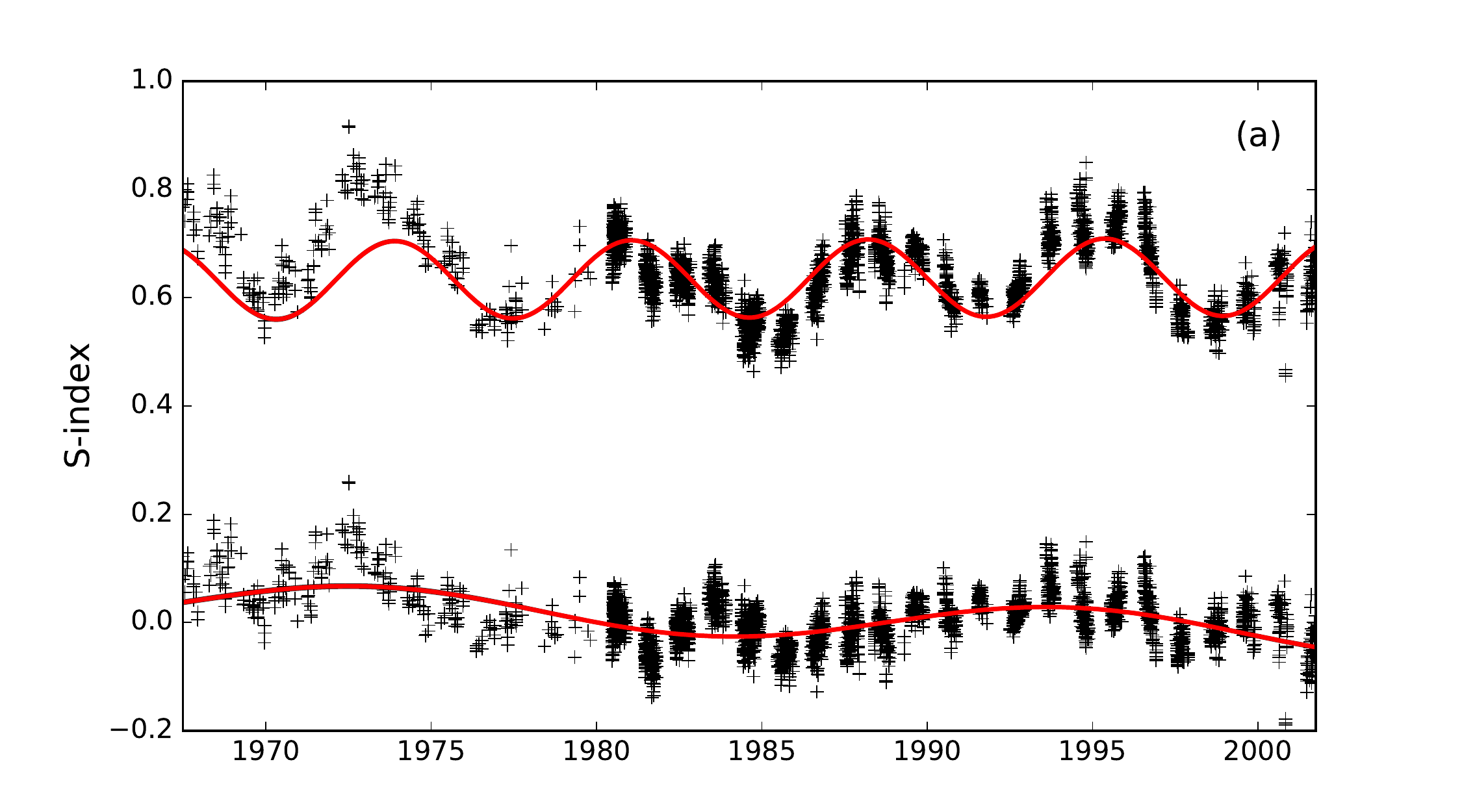} \\
		\includegraphics[width=0.5\textwidth, trim={0 0.6cm 0 1cm}, clip]{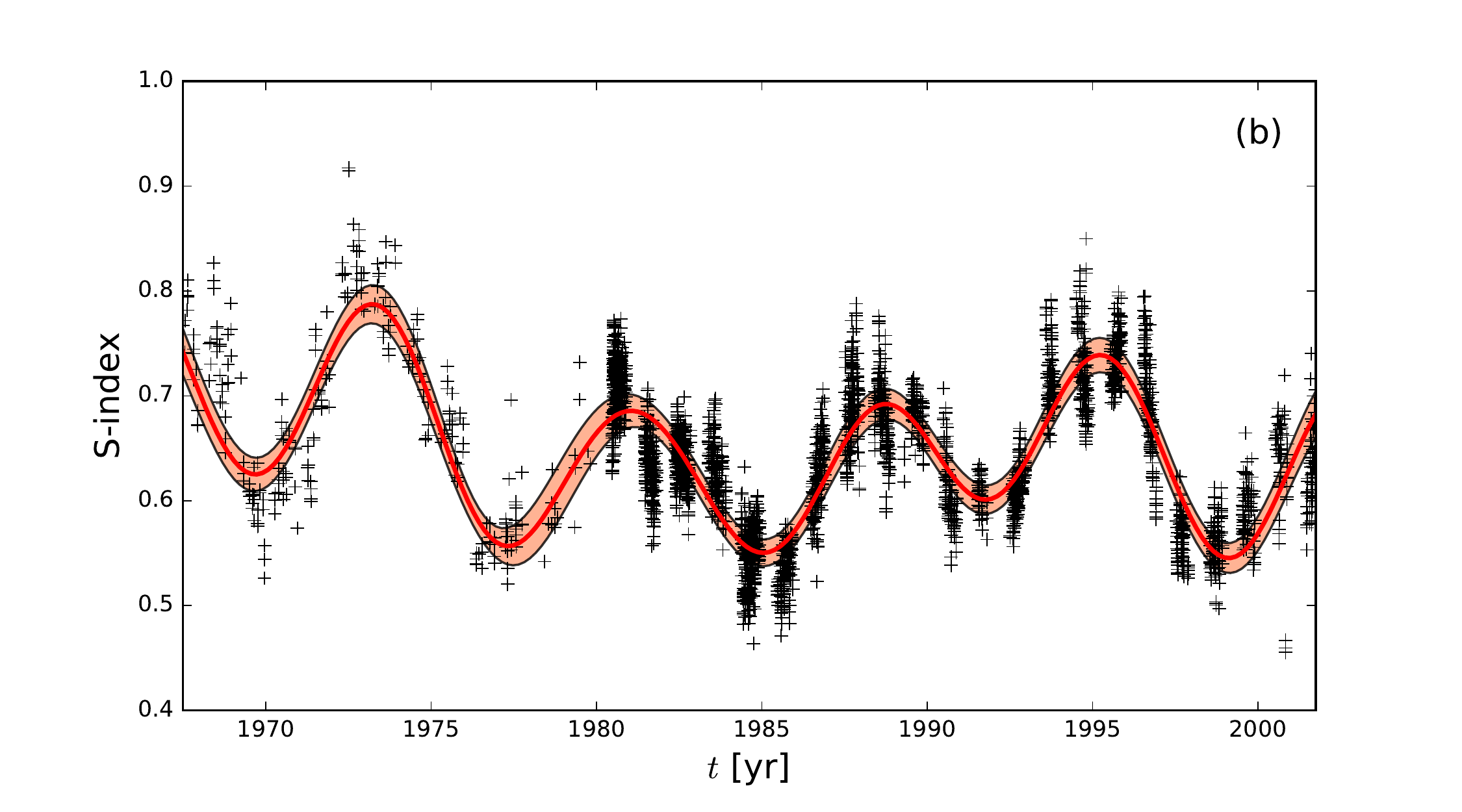} \\
		\includegraphics[width=0.5\textwidth, trim={0 0cm 0 1cm}, clip]{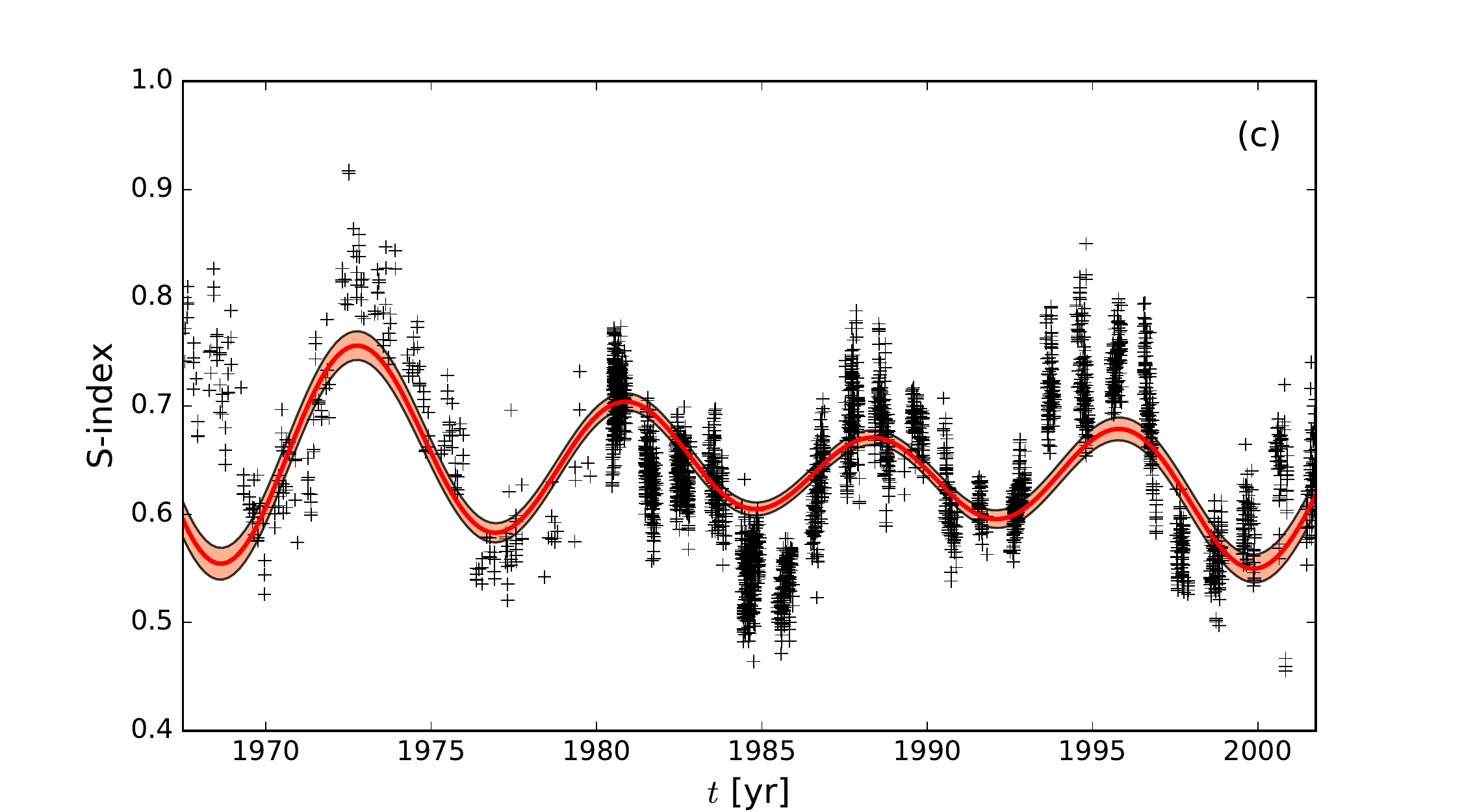}
	\end{tabular}
	\caption{Comparison of the fits of H (a), P (b) and QP (c) 
		models for HD201091. The meaning of the panels and symbols is the same as in Fig.~\ref{fig_sun_fits}.}
	\label{fig_201091_fits}
\end{figure}

As a last example we want to show why the P model turned out to be quite impractical in some of the cases, 
because preferring longer periods with less harmonic phase behaviour over shorter and more harmonic periods. 
For instance for the
star HD201091 there is obviously a clear cycle present in the data around 7-8 yrs. However, the optimal fit from the
P corresponds to a very long period around 29 yrs, which is almost the length of the dataset, 
thus being neglected from the results. Neither is this complex phase behaviour physically meaningful. 
To avoid such fits we could have used more restrictive priors on $\ell$, but we decided not to do it to allow finding
better fits than those obtained already by using H model.
The comparison of the fits for this star are shown in Fig.~\ref{fig_201091_fits}. On panel (a) we show both
of the cycles detected by the harmonic model plotted on top of the original data and the residuals correspondingly.
The time-scale of the QP model in this case is approximately 14 times longer than the cycle period.

\end{appendix}
\end{document}